\documentclass[prx,twocolumn,english,longbibliography,superscriptaddress]{revtex4-2}
\usepackage{ifthen}
\usepackage{xcolor}
\usepackage{graphics,graphicx,epsfig}
\usepackage{epsf,epstopdf,wrapfig}
\usepackage{amssymb,amsfonts,amsmath}
\usepackage[export]{adjustbox}
\usepackage[english]{babel}
\include{epsf}
\usepackage{ifthen}
\usepackage{algorithm2e}
\usepackage{rotating} 

\definecolor{ocre}{RGB}{10,100,185}
\usepackage[colorlinks = true,
            linkcolor = ocre,
            urlcolor  = ocre,
            citecolor = ocre,
            anchorcolor = ocre]{hyperref}

\usepackage[normalem]{ulem}
\usepackage{booktabs}
\usepackage{multirow}

\usepackage{makecell}

\newcommand{\EQ}{\begin{equation}}
\newcommand{\EE}{\end{equation}}
\newcommand{\EQA}{\begin{eqnarray}}
\newcommand{\EEA}{\end{eqnarray}}

\newcommand{\unibeta}{B}

\newcommand{\trb}{TR{\unibeta}}
\newcommand{\pgen}{P_{\mathrm{gen}}}
\newcommand{\ppost}{P_{\mathrm{post}}}
\def\nt{\ensuremath \mathrm{nt}}

\def\tcrdist{\ensuremath \mathrm{TCRdist}}

\newcommand{\tcrb}{TCR{$\beta$}}
\newcommand{\pobs}{P_{\mathrm{obs}}}

\newcommand{\pshare}{P_{\mathrm{share}}}

\newcommand{\RNum}[1]{\uppercase\expandafter{\romannumeral #1\relax}}

\newcommand{\corres}{Correspondence should be addressed to:
 Armita Nourmohammad: {armita.nourmohammad@yale.edu}.}

\begin{document}

\title{T-cell repertoire response in individuals with post-acute sequelae of COVID-19}
\author{Zachary Montague}
\affiliation{Department of Physics, University of Washington, 3910 15th Avenue Northeast, Seattle, WA 98195, USA}
\author{Rhea  M Grover}
\affiliation{Department of Physics, University of Washington, 3910 15th Avenue Northeast, Seattle, WA 98195, USA}
\affiliation{Department of Mathematics, University of California, Berkeley, Berkeley, CA, USA}
\affiliation{Department of Applied Mathematics, University of Washington, 4182 W Stevens Way NE, Seattle, WA 98105, USA}
\author{Andrew Baumgartner}
\affiliation{Institute for Systems Biology, Seattle, WA, USA}
\affiliation{Phenome Health, 2101 4th Ave, Seattle, WA 98121, USA}
\author{Assya Trofimov}
 \affiliation{Department of Physics, University of Washington, 3910 15th Avenue Northeast, Seattle, WA 98195, USA}
 \affiliation{Fred Hutchinson Cancer Center, 1241 Eastlake Ave E, Seattle, WA 98102, USA}
\author{Jennifer Hadlock}
\affiliation{Institute for Systems Biology, Seattle, WA, USA}
\affiliation{Department of Biomedical Informatics and Medical Education, University of Washington, Seattle, WA, USA}
\author{Armita Nourmohammad}
\thanks{\corres}
 \affiliation{Department of Physics, University of Washington, 3910 15th Avenue Northeast, Seattle, WA 98195, USA}
 \affiliation{Department of Applied Mathematics, University of Washington, 4182 W Stevens Way NE, Seattle, WA 98105, USA}
\affiliation{Fred Hutchinson Cancer Center, 1241 Eastlake Ave E, Seattle, WA 98102, USA}
\affiliation{Paul G. Allen School of Computer Science and Engineering, University of Washington, 85 E Stevens Way NE, Seattle, WA 98195, USA}
\affiliation{Center for Systems and Engineering Immunology, and the Departments of Immunobiology, Biomedical Engineering, and Physics,  Yale University, New Haven, CT, USA}

\begin{abstract}
\noindent
T-cells are central to SARS-CoV-2 clearance and immunological memory, yet their contribution to the persistence of post-acute sequelae of COVID-19 (PASC) remains poorly understood. The immunological features that distinguish individuals who develop PASC from those who recover fully are unresolved, in part due to the phenotypic heterogeneity of the condition and the likely multiplicity of its underlying mechanisms. Here, we profiled longitudinal bulk TCR$\beta$ repertoires from 120 individuals in the INCOV cohort--71 with PASC and 49 without--sampled at two to three time points spanning the acute and post-acute phases of infection. Using robust statistical modeling of repertoire composition and clonal dynamics, we found that global statistics such as V, J gene usage and CDR3 length do not differ between groups, but that locally enriched sequence motifs and differentially dynamic clones reveal distinct T-cell signatures associated with PASC status. Clones contracting following the peak of the acute response were significantly enriched for SARS-CoV-2 specificity in both groups. Interestingly, Influenza A-specific TCRs were disproportionately enriched among contracting  clones in PASC{$^+$} repertoires, implicating viral co-infection as a potential contributor to early disease severity and, possibly, PASC pathogenesis. Rare public TCR clones were markedly enriched for SARS-CoV-2 specificity, with PASC{$^+$} individuals harboring a modestly but significantly higher proportion than PASC{$^-$} individuals. Together, we identified over 1,000 candidate TCR$\beta$ receptors potentially discriminating PASC{$^+$} from PASC{$^-$} immune responses, opening a path toward the identification of disease-relevant T-cell specificities and the development of T-cell-based immunological biomarkers for long COVID.
\end{abstract}
\maketitle

\section{Introduction} 
Over five years since the inception of the COVID-19 pandemic, millions of individuals afflicted previously with SARS-CoV-2 infections have persisting maladies, a condition called post-acute sequelae of SARS-CoV-2 infection (PASC) and, colloquially, Long COVID~\cite{davis2023long,greenhalgh2024long,michelen2021characterising,parotto2023post,jiang2021postacute,devries2023one}. 
PASC is characterized by continuous, relapsing and remitting, or progressive disease state that affects one or more organ systems, and can include a wide range of conditions, including fatigue, neurocognitive dysfunction, hyperglycemia, cardiovascular disease, autonomic dysfunction, and gastrointestinal disorders~\cite{mancini2021use,kedor2022prospective,palmer2024non,tsampasian2024cardiovascular,xie2022long,kedor2022prospective,al2024long,pollack2023female,xu2023long,National-Academies-of-Sciences-Engineering-and-Medicine;-Health-and-Medicine-Division;-Board-on-Global-Health;-Board-on-Health-Sciences-Policy;-Committee-on-Examining-the-Working-Definition-for-Long-COVID2024-ip,Geng2025-oo}.
The etiologies underlying some of these symptoms are believed to be uncleared viral reservoirs~\cite{swank2022persistent}, autoimmune responses~\cite{su2022multiple}, chronic inflammation~\cite{davis2023long}, and the reactivation of quiescent viruses~\cite{zubchenko2022herpesvirus,wong2023serotonin,klein2023distinguishing,peluso2023chronic} among others~\cite{davis2023long,mohandas2023immune,yin2024long}.
PASC can present in individuals regardless of age, sex, or COVID-19 severity, though the likelihood of its presentation is associated positively with the severity an individual's COVID-19~\cite{cai2024three,Wuller2025-rl}. While methods for preventing PASC mirror those for SARS-CoV-2, significant breakthroughs in diagnosing and treating PASC are hampered by the wide heterogeneity of the disease and unclear pathogenesis~\cite{al2024solving,Peluso2024-md}.

\begin{figure*}[th!]
\includegraphics[width=0.95\textwidth]{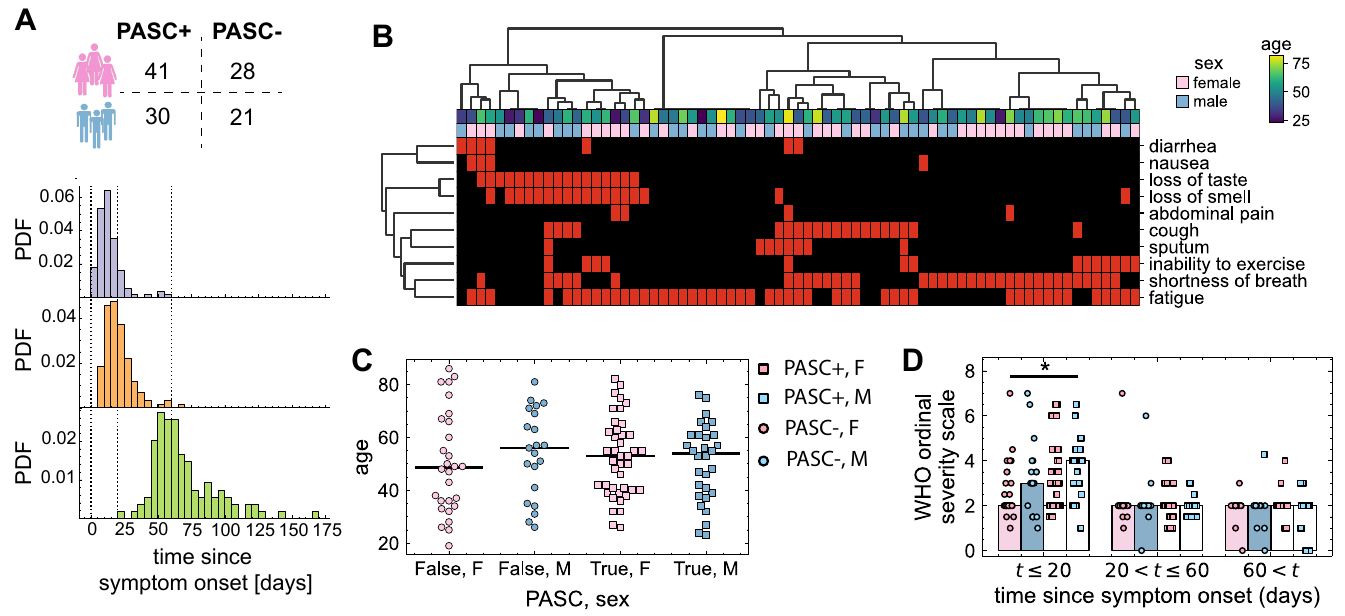}
  \caption{
  {\bf PASC symptoms, sampling time distributions, and cohort demographics.} 
   {\bf(A)} Separation of individuals into groups based on their PASC status and sex is shown. The histograms show the distributions of the times at which repertoires were sampled relative to an individual's onset of symptoms. From top to bottom: earliest sampled time, second time point, and third sampled time point. 
The dotted vertical lines are drawn at x values of $0$, $20$, and $60$ to indicate how samples are binned.
   {\bf(B)} PASC symptoms (y-axis) presented by individuals in the PASC{$^+$} group (x-axis) are shown as a clustermap created using Ward clustering by Jaccard distances between sets of symptoms.   
  Red rectangles display which symptoms were observed.
  The left dendrogram shows the relationships among symptom presentation whereas the top dendrogram shows clusters of individuals presenting similar symptoms.
  The colored rectangles directly above the heatmap show the sex (blue for male, pink for female) of each individual in the PASC{$^+$} group, and the viridis-colored rectangles show the age of each individual in the PASC{$^+$} group, with the associated colorbar displayed  on the left. 
   {\bf(C)} The swarmplot shows the distribution of ages of the PASC{$^+$} and PASC{$^-$} groups stratified by sex (x-axis, colors, marker shapes).
  Horizontal lines show the median age of each swarm.
  {\bf(D)} The WHO ordinal severity scale (WOSS) measures the disease severity of an individual via their respiratory status~\cite{marshall2020minimal}. Bars show the median WOSS value (y-axis) for each group partitioned by PASC status (hatches/marker shapes as in (C)) and sex (colors as in (C)) for each binning of observation times (x-axis). 
  *:~Bonferroni-corrected $p$-value $< 0.05$.
      \label{fig:tcrcovid_fig1}
 }
 \end{figure*}
T-cells are crucial for surveillance and clearing infections and for protection against previously encountered pathogens through memory~\cite{janeway2016immunobiology,lam2024guide,maltzman2023cd8,kunzli2023cd4+}. T-cell dysregulation has been proposed as one of the etiologies of PASC~\cite{yin2024long}, and the resolution of dysregulation in the adaptive immune system has been associated with an improvement in the quality of life of individuals with PASC~\cite{phetsouphanh2024improvement}.
 Machine learning can identify distinct T-cell subsets associated with the severity of restrictive lung disease and breathlessness in PASC ~\cite{canderan2025distinct}.
Because leading PASC hypotheses--persistent viral reservoirs, chronic inflammation, or both--implicate T‑cell activity, dissecting the SARS‑CoV‑2‑specific T‑cell response is pivotal for clarifying disease mechanisms and guiding therapy design.

T-cells present a diverse repertoire of antigen-engaging T-cell receptors (TCRs), which are generated in a stochastic process called V(D)J recombination.  
Investigating T-cell repertoires has elucidated how the adaptive immune system responds to various pathogens and challenges~\cite{emerson2017immunosequencing,pogorelyy2024tirtl,minervina2020primary,minervina2021longitudinal,rawat2024identification,pogorelyy2018method,pogorelyy2019detecting,vujkovic2024diagnosing,gittelman2022longitudinal}. Although many studies have profiled T‑cell responses in PASC~\cite{santa2023post,su2022multiple,
klein2023distinguishing,littlefield2022sars,williams2024investigating,paniskaki2023increased,yin2024long,phetsouphanh2024improvement}, none, to our knowledge, 
 has mapped the disorder's imprint on patients' bulk TCR repertoires. Longitudinal, deep‑sequencing of these repertoires can track millions of clonotypes over time, revealing clonal expansions, contractions and shared antigen‑driven motifs. Understanding  T-cell responses in the context of PASC and identifying commonalities in TCRs observed in individuals with PASC could   provide new insight into disease mechanisms. 

This study presents a comprehensive T-cell receptor ({$\beta$} chain) repertoire dataset  derived from peripheral blood samples of 120 participants in the INCOV cohort with confirmed SARS-CoV-2 infection, comprising 49 individuals who were admitted to the hospital for acute COVID-19, and 71 with post-acute sequelae of COVID-19 (PASC). Samples were collected longitudinally at two to three time points following symptom onset, enabling temporal analysis of immune responses. We characterize the statistics and dynamics of T-cell repertoires in PASC  and non-PASC patients. Through robust statistical modeling, we identify and probe TCRs of interest such as those with large clones, public clones, and  those that significantly expanded or contracted following infections, and characterize the differential T-cell repertoire features between  PASC  and non-PASC patients. Ultimately, we identify 1,091 TCR $\beta$-chain ({\trb}) receptors that are candidates for potential association with PASC{$^+$} or PASC{$^-$} responses. 

\section{Overview of the cohort and clinical data}
We analyzed bulk T-cell repertoires of 120 participants from the INCOV cohort, which was first established to profile acute SARS‑CoV‑2 infections~\cite{su2020multi} and later expanded to investigate post‑acute sequelae of COVID‑19 (PASC)~\cite{su2022multiple}. Peripheral blood mononuclear cells were collected from each participant and bulk T-cell receptor {\trb} sequencing was carried out by Adaptive Biotechnologies (Methods). Longitudinal sampling captured two to three time points per individual, ranging from 2 to 169 days after enrollment, which approximately corresponds to the patients' symptom onset; we refer to this time point as  symptom onset for simplicity  (Fig.~\ref{fig:tcrcovid_fig1}A).

A uniform case definition for post‑acute sequelae of COVID‑19 (PASC) is still lacking~\cite{danesh2023symptom,reese2023generalisable,hanson2022estimated,al2024solving,National-Academies-of-Sciences-Engineering-and-Medicine;-Health-and-Medicine-Division;-Board-on-Global-Health;-Board-on-Health-Sciences-Policy;-Committee-on-Examining-the-Working-Definition-for-Long-COVID2024-ip,Geng2025-oo}. In practice, however ($>3$ months) of one or more of the following symptoms --- shortness of breath, exercise intolerance, cough, sputum production, fatigue, diarrhea, nausea, abdominal pain, loss of taste, or loss of smell --- are accepted as indicative of PASC. 

These manifestations span multiple organ systems, and their etiologies are likely heterogeneous~\cite{davis2023long}. Within our cohort, the symptoms clustered into five partially overlapping groups: diarrhea and nausea (gastrointestinal); loss of taste and smell (anosmia/dysgeusia); abdominal pain; cough and excretion of sputum (respiratory); and fatigue, shortness of breath, and inability to exercise (neurological); Fig.~\ref{fig:tcrcovid_fig1}B. Because of the high degree of co‑occurrence and the modest sample size, in our analyses we classified participants as PASC{$^+$} if they reported at least one symptom within one of these clusters after two months post infection. This yielded 71 PASC{$^+$}  individuals  (58\% female, median age 53) and 49 PASC{$^-$} individuals (57\% female, median age 50) (Fig.~\ref{fig:tcrcovid_fig1}A). Symptom prevalence did not differ significantly by age or sex (Methods; Fig.\ref{fig:tcrcovid_fig1}B). It is worth noting that as our understanding of PASC and its disease subtypes evolves, possible refinements to this stratification may have implications for some of our conclusions.

Ages of the participants spanned over 19-86 years with comparable distributions across PASC status and sex strata (Fig.~\ref{fig:tcrcovid_fig1}C). All infections occurred during the COVID-19 pandemic's first year (year 2020) and covered the full clinical spectrum, assessed with the WHO Ordinal Severity Scale (WOSS)~\cite{marshall2020minimal}. Fig.~\ref{fig:tcrcovid_fig1}D summarizes WOSS scores in three windows: $<20$ days, 20 - 60 days, and $>60$ days post‑symptom onset. The only significant difference is shown in the early time window ($<20$ days post‑symptom onset), during which  PASC{$^-$}  females show more severe symptoms (higher  WOSS scores) compared to  PASC{$^+$} males  (Mann-Whitney $U=469$, $n_{-,F} = n_{+,M} = 25$, Bonferroni-corrected $\text{p-value}= 0.036$).  Consistent with prior observations~\cite{keur2024delineating,sieurin2022population}, males showed marginally greater severity within the first three weeks of infection, but median severity was mild across all PASC and sex groups thereafter.

 \begin{figure*}[htbp]
{\includegraphics[width=1\textwidth]{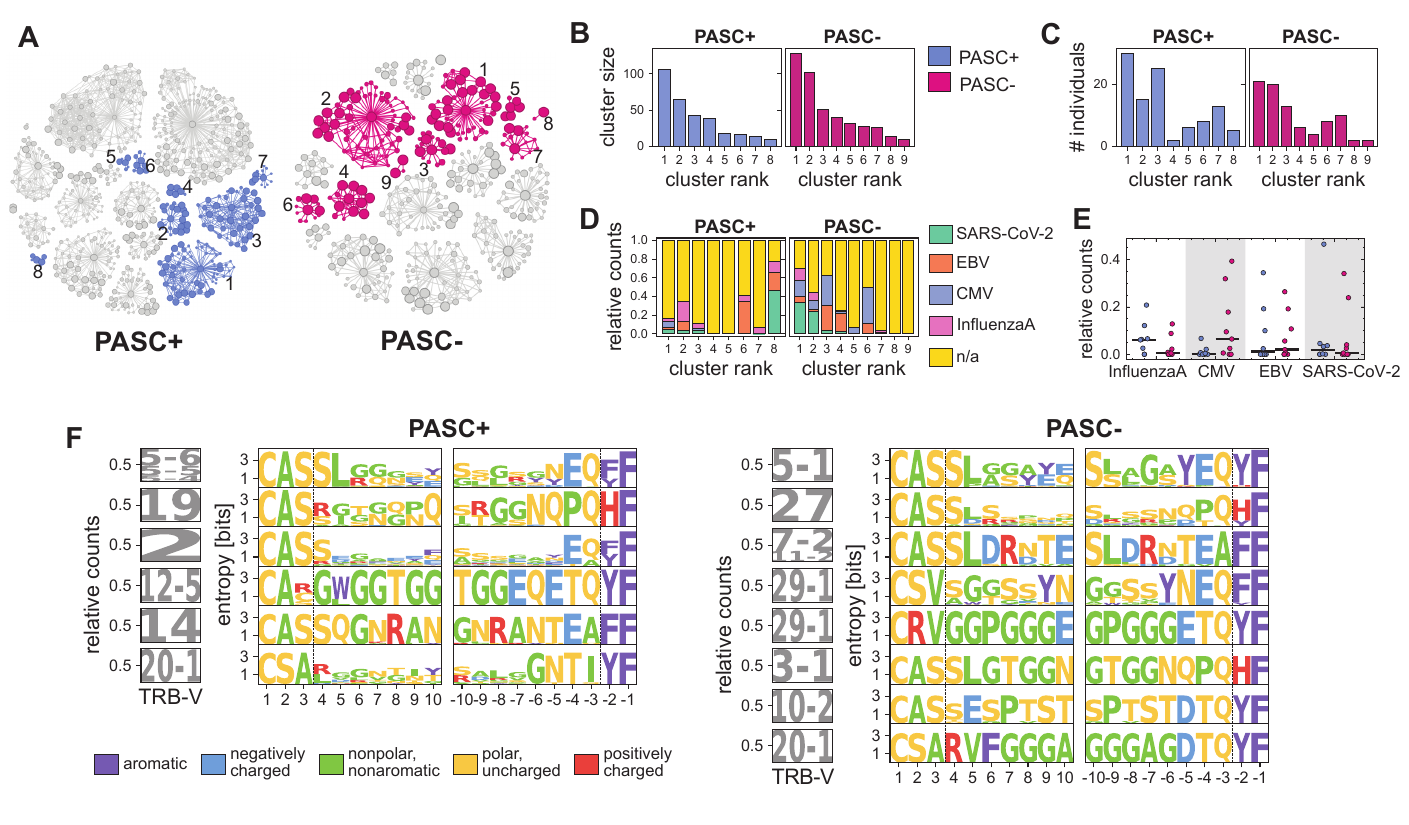}}
    \caption[Top rank-50 receptor composition and TCR-OT clusters.]{
    {\bf Top rank-50 receptor composition and TCR-OT clusters.}
    {\bf(A)} The network plots show the topology of the TCR-OT clusters for the PASC{$^+$} (left) and PASC{$^-$} (right) patients.
    Sizes of the nodes show the strength of the $-\log_{10}$ FDR-BH corrected $p$-values associated with a functional clone, with larger sizes indicating more significantly atypical functional clones.
    Edges between nodes and the cluster's focal sequence indicate sequence similarity within the TCR-OT inferred cluster radius, and all other edges indicate sequence similarity within 48 TCRdist~\cite{dash2017quantifiable}; this threshold amounts to a difference of about 4 amino acids in the CDR3 sequences or 12 amino acid differences between the V genes of two TCR$\beta$ sequence.
    Network topologies with colored nodes had at least 20\% of their nodes deemed significant and had recombinations from at least two individuals.
    The numbers next to each colored cluster indicate the ranking of the clusters with respect to the cluster sizes, with 1 indicating the cluster with the largest amount of clones.
      {\bf(B)} The amount of independent recombinations in the clusters for each group (colors) is shown as a bar graph. 
    Cluster rank corresponds to the size of the clustering.
    {\bf(C)} The number of individuals represented in a cluster for each group (colors similar to B) is shown as a bar graph.
         {\bf(D)} The stacked bar graphs show the distribution of epitopes of {\trb} sequences in VDJdb that were within 24 TCRdist of a constituent {\trb} in a cluster and whose donor HLA group matched to at least one HLA group of the individual from which the similar {\trb} sequence originated.
    The left bar graph shows the distribution for the PASC{$^+$} clusters while the right bar graph shows the distribution for the PASC{$^-$} clusters.
    {\bf(E)} The epitope distributions of (D) are shown as a strip plot for the PASC{$^+$} and PASC{$^-$} patients (colors), with the horizontal black lines indicating the medians of the distributions (colors similar to B).
     {\bf(F)} Sequence logos describing the composition of amino acid sequences are shown for the significant TCR-OT clusters obtained from the PASC{$^+$} (left) and  PASC{$^-$}  (right) patients.
    Logos were constructed from focal sequence(s) and members in the cluster that were within 48 TCRdist of the focal sequence(s).
    Logos are ordered top to bottom, with the top logo describing the cluster with the largest size and the bottom logo describing the cluster with the smallest size.
    Only clusters with at least 10 clones within a neighborhood of 48 TCRdist of the focal sequence(s) are plotted.
     V gene usage frequencies within a cluster are shown as a logo  (Left). For CDR3 amino acid composition (right), the logo plots shown the information content at each position in the   CDR3 sequence from left to right (left) and right to left (right).  Amino acids are colored according to their physicochemical properties indicated in the legend.
     The first three positions and last two positions of the amino acid sequence do not enter into TCRdist calculations (Methods), and a vertical, dashed line demarcates where this occurs in each respective logo.
    \label{fig:tcrcovid_tcrot_composition}
    }
\end{figure*}

 \section{Impact of PASC  on the statistics of TCR repertoires}

\noindent{\bf Broad TCR repertoire statistics do not differ between PASC$^+$ and PASC$^-$ patients.} 
We sought to determine how PASC status shapes the statistics of TCR sequences, such as   V gene and J gene usages, and CDR3 amino acid sequence lengths. These sequence features, while also associated with an individual's HLAs, can contain information pointing to a receptor's binding specificity.  Overall, we did not see statistically significant differences in gene usage and CDR3 lengths of TCRs between PASC{$^+$} and PASC{$^-$} patients  (Fig.~\ref{fig:SI:tcrcovid_composition}A-D, Tables~\ref{table:vgene_ttests}--\ref{table: cdr3_ttests}). However, compared to healthy repertoires, both groups show slightly (but significantly) shorter CDR3 sequences (Table~\ref{table: cdr3_ttests}). Moreover,  in both groups we observe an increased usage of TRBV20-1, V3, and V9,  and decreased usage of TRBV10, V2, V5, and V6 genes, consistent with prior findings~\cite{Hou2021-xn, Luo2021-cs} (Table~\ref{table:vgene_ttests}).

Clonal composition of repertoires can change significantly in response to immune challenges, as perhaps most dramatically observed in CMV patients, where their repertoires can be dominated by  a few large clones~\cite{Weekes1999-re}. However, comparing the clonality of PASC{$^+$} and PASC{$^-$} individuals, we found no significant differences between their clonal frequency distributions,  across sampled time points and sexes (Methods and Fig. \ref{fig:SI:tcrcovid_composition}E). This suggests that PASC does not modulate the distribution of clone sizes in a manner distinct from acute COVID-19.
Overall PASC status does not seem to influence the global sequence statistics and clonal composition of individuals' TCR repertoires (Fig.~\ref{fig:SI:tcrcovid_composition}).\\

\noindent{\bf PASC$^-$specific TCR motif enrichments.}
Despite the absence of  global differences in receptor repertoire statistics between PASC{$^+$} and PASC{$^-$} patients (Figs.~\ref{fig:SI:tcrcovid_composition}A-D,\ref{fig:tcrcovid_tcrot_top50_clusterfit}), we hypothesized that functionally relevant signals may be confined to restricted regions of sequence space. To extract such local enrichments, we used TCR‑OT~\cite{olson2022comparing}, a non‑parametric optimal‑transport framework that treats each repertoire as a probability distribution over deduplicated   V-gene and CDR3 amino‑acid sequences. TCR‑OT derives an entropy‑regularized {\em transport plan} between two repertoires, yielding a Wasserstein distance between repertoires (Methods). Beyond this global metric, TCR‑OT assigns an enrichment score  to every sequence in a query repertoire relative to a reference  repertoire. The significantly enriched clones can be agglomerated into clusters (defining sequence motifs), which can reflect  sequence signatures of  shared, antigen‑driven functionality of the query set that is distinct from that of the reference repertoire; see Methods for details on the approach and algorithm.

To compare differential motif enrichments between PASC{$^+$} and PASC{$^-$} patients, we first focused on the 50 most abundant clones within each donor, defined using  CDR3 amino acid sequences and V genes (Methods). 
We restricted this analysis to the 62 individuals observed at least 60 days after enrollment in the study. This criteria assures that our analysis would not be dominated by clones not observed well after the onset of PASC in PASC{$^+$} individuals or convalescence in PASC{$^-$} individuals.
These filtering criteria resulted in 2,628 clones (3,357 independent CDR3 recombinations) from the PASC{$^+$} patients and 1,535  clones (2,019 independent CDR3 recombinations) from the PASC{$^-$} patients.

While the gene usage and the CDR3 length statistics are similar between groups (Fig.~\ref{fig:tcrcovid_tcrot_top50_clusterfit}A-C), we detected 159  clones in the PASC{$^+$} and 105  clones in the PASC{$^-$} patients that were significantly atypical in amino acid composition relative to the other patient  group (FDR BH-corrected $p$-value $< 0.05$); Fig.~\ref{fig:SI:tcrcovid_tcrot_top50_composition}D,E. 

Seeding the TCR‑OT clustering algorithm with the most enriched clones produced 20 clusters per group  (Fig.~\ref{fig:tcrcovid_tcrot_top50_clusterfit}), whose sizes ranged from 2-147 clones
in  the PASC{$^+$} group, and 1-57 clones 
in the PASC{$^-$} group (Fig. \ref{fig:tcrcovid_tcrot_composition}A). Because the enrichment significances of the sequences varied within clusters (Fig.~\ref{fig:SI:tcrcovid_tcrot_top50_composition}D) and to ensure that clusters reflected a degree of collective neighborhood amplification associated with PASC status of a group, we required that (i) at least 20\% of the clones in a cluster to be significantly enriched relative to the other group, and (ii) that TCRs from at least two individuals to be present in each TCR-OT cluster. These clusters show a broad range of sizes and number of individuals represented in them (Fig. \ref{fig:tcrcovid_tcrot_composition}B,C). 

To infer possible antigenic drivers, we matched sequences in the identified TCR-OT clusters to sequences previously reported as specific to SARS-CoV-2, Epstein-Barr Virus (EBV), Cytomegalovirus (CMV), or Influenza A in the VDJdb~\cite{shugay2018vdjdb,goncharov2022vdjdb}, accepting matches within 24 TCRdist~\cite{dash2017quantifiable} 
 (corresponding to TCRs with $\approx 2$ CDR3 substitutions and identical TRBV), and at least one shared HLA allele (Fig.~\ref{fig:tcrcovid_tcrot_composition}D). 
We analyzed TCR sequences in our repertoires for these pathogens since EBV~\cite{gold2021investigation,klein2023distinguishing,rohrhofer2022association,su2022multiple} and CMV~\cite{cervia2024persistent} viral reservoirs have been hypothesized to be reactivated in PASC{$^+$} individuals, though studies are not completely concordant in these observations~\cite{su2022multiple,hoeggerl2023epstein,cervia2024persistent}.  Of the sequences in the PASC{$^+$} clusters that matched to VDJdb, we observed 1--7 matches to CMV, 1--4 matches to Influenza A, 1--6 matches to SARS-CoV-2, and 2--4 matches to EBV; of the sequences in the PASC{$^-$} clusters that matched to VDJdb, we observed 1--14 matches to CMV, 1--8 matches to InfluenzaA, 1--11 matches to SARS-CoV-2, and 1--7 matches to EBV (Fig.~\ref{fig:tcrcovid_tcrot_composition}D).  While {\trb} sequences with known Influenza A epitopes were slightly enriched in PASC{$^+$} clusters and those with CMV epitopes were slightly enriched in PASC{$^-$} clusters, these enrichments were not statistically significant (Mann-Whitney U test, Bonferroni-corrected $p > 0.5$) (Fig.~\ref{fig:tcrcovid_tcrot_composition}E).

Lastly, we characterized the sequence motifs associated with each of the identified  clusters in Fig.~\ref{fig:tcrcovid_tcrot_composition}F.  Sequence logos constructed from neighborhoods with a radius of 48 TCRdist~\cite{dash2017quantifiable}  around the focal clones of significant TCR‑OT clusters highlighted recurrent use of TRBV5‑5 and TRBV20‑1 in PASC{$^+$} and TRBV5‑1, TRBV20‑1 and TRBV29‑1 in PASC{$^-$} patients, mirroring TRBV genes previously linked to SARS‑CoV‑2 reactivity~\cite{pogorelyy2022resolving}.

\begin{figure*}[htbp]
\includegraphics[width=0.95\textwidth]{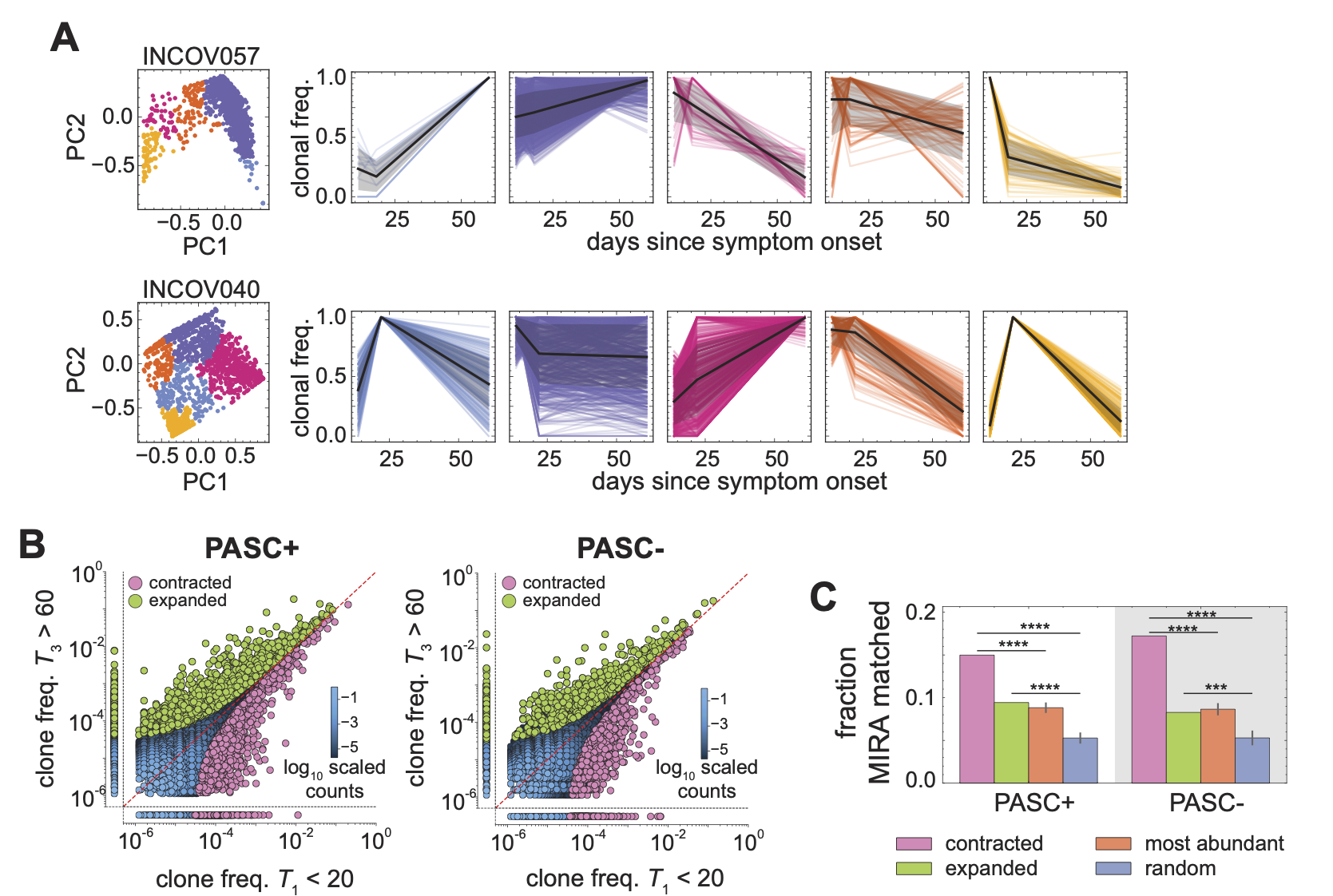}
\caption[Dynamics of TCR repertoires.]{
    {\bf Dynamics of TCR repertoires.}
    {\bf(A)} Dynamical modes of the TCR repertoire of in two individuals (INCOV057 and INCOV040).
  (left) PCA of normalized top rank-1000 clonal frequencies.
  Points are colored by the output of hierarchical clustering.
  (right) The normalized clonal trajectories of clones within a cluster (colors as in Left) are plotted as lines.
  The black line shows the average clonal trajectory, and shading indicates regions containing one standard deviation of variation across trajectories in that mode.
  {\bf(B)} The normalized clonal frequencies observed in the earlier sampled repertoire (x-axis) and later sampled repertoire (y-axis) are shown as a scatter plot for clones pooled from the PASC{$^+$} patients (left) and clones pooled from the PASC{$^-$} patients (right).
  In each panel, regions bounded by black, dashed lines show clones present in only the early sample (bottom) or later sample (left).
  Lighter colors indicate higher densities of points.
  Clones detected as significantly contracted are colored pink, and clones detected as significantly expanded are colored green.
  {\bf(C)} The average fraction of clones which had a match to MIRA (i.e., was a neighbor to at least two MIRA sequences within 12 TCRdist) for the dynamically significant clones (pink, green) are shown for PASC{$^+$}(left) and PASC{$^-$} (right) patients. Controls indicate MIRA-overlap with 250 subsamples of top rank-200 clones (orange), and 250 subsamples of randomly chosen clones (purple) from the respective dataset (x-axis). 
  ****: $p$-value $< 10^{-4}$; ***: $p$-value $< 10^{-3}$.
  \label{fig:clonal_expansion}
}
\end{figure*}

\begin{figure*}[htbp]
\includegraphics[width=0.95\textwidth]{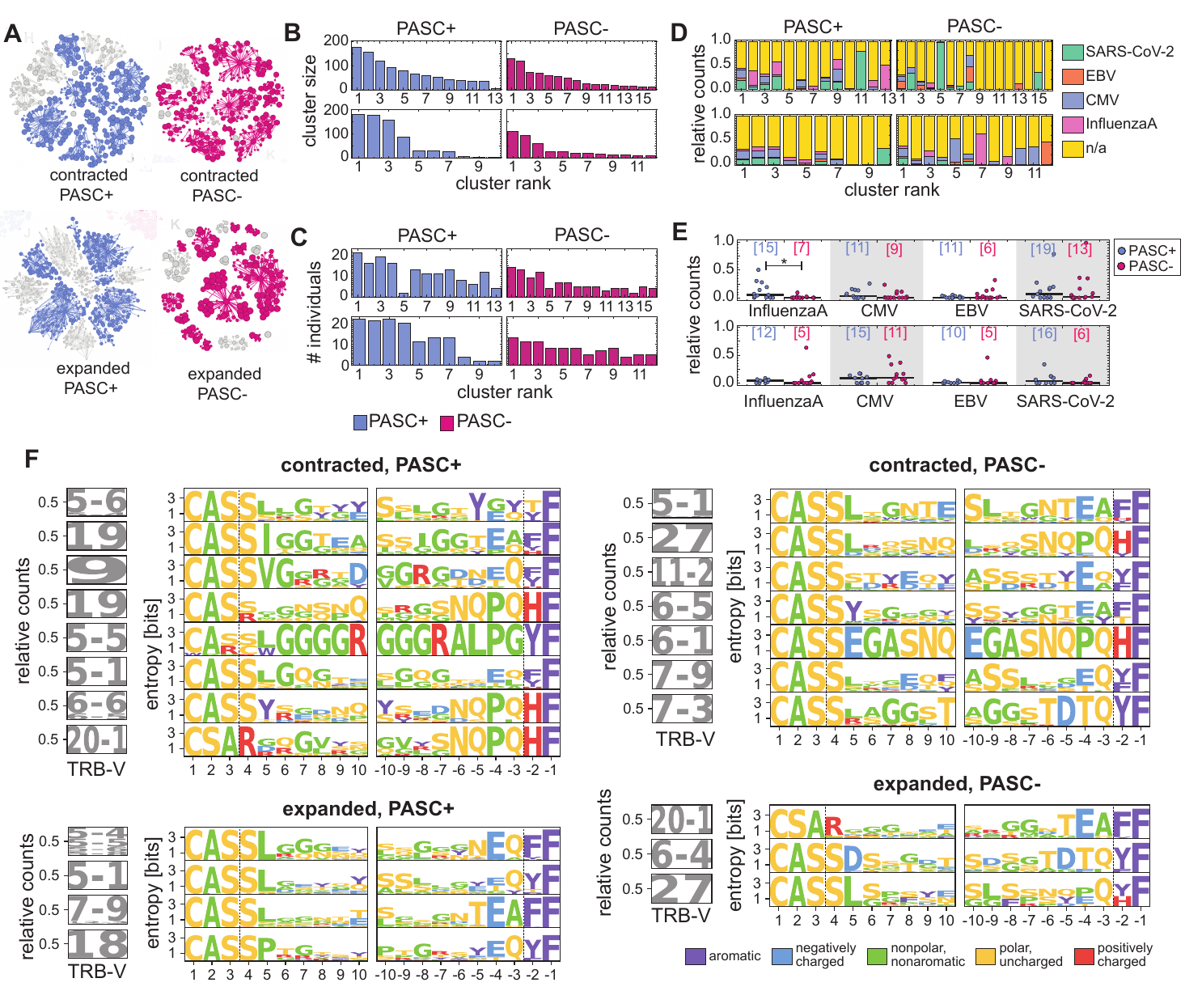}
\caption{
    {\bf Differential features of dynamic TCR repertoires between PASC{$^+$} and PASC{$^-$} patients.}
    {\bf (A)} The network plots show the topology of the TCR-OT clusters for the PASC{$^+$} / PASC{$^-$} and   contracted / expanded subsets. Network annotation and coloring scheme similar to Fig.~\ref{fig:tcrcovid_tcrot_composition}. {\bf (B)} The number of independent recombinations and   {\bf (C)} the number of individuals represented in the TCR-OT clusters  from the contracted (top) and expanded (bottom) repertoires of PASC{$^+$}  and PASC{$^-$} individuals (colors) are shown as bar graphs. Cluster rank corresponds to the size of the cluster, {with the largest cluster being the highest rank}. {\bf (D)} The stacked bar graphs show
the distribution of epitopes of TCR sequences in VDJdb that were within 24 TCRdist of a constituent TCR in a cluster
and whose donor HLA group matched to at least one HLA group of the individual from which the similar  TCR sequence was originated (Methods). Panels show cluster statistics for  contracted (top), expanded (bottom) sub-repertoires collected from  PASC{$^+$} (left),  and  PASC{$^-$} (right) individuals.
 { {\bf (E)} The epitope distributions of (D) are shown as a strip plot for the  contracted (top) or expanded (bottom) TCR-OT clusters from PASC{$^+$}  and PASC{$^-$}   
 The horizontal black lines show the medians. $\star$: Bonferroni-corrected $p$-value $<0.05$. Numbers indicate the number of individuals associated with sequences in each subset.  {\bf (F)}
Sequence logos describing the composition of amino acid sequences are shown for the significant TCR-OT clusters obtained from the contracted / expanded  PASC{$^+$} and PASC{$^-$} subsets; similar convention to Fig.~\ref{fig:tcrcovid_tcrot_composition}F is used.  Logos were constructed from focal sequence(s) and members in the cluster that were within 48 TCRdist of the focal sequence(s). Logos are shown only if 50 sequences were present in this 48 TCRdist neighborhood and are ordered top to bottom by cluster size. {Logo plots for smaller clusters are shown in} Fig.~\ref{fig:SI_tcrcovid_tcrot_dynamics_motifs}. 
} 
  \label{fig:clonal_expansion_OT}
}
\end{figure*}

\section{Impact of PASC on the TCR repertoire dynamics}

\noindent{\bf Dynamical modes of TCR repertoires in PASC{$^+$} and PASC{$^-$} patients.} T‑cell responses typically peak 10–14 days after antigen encounter~\cite{miller2008human,pogorelyy2018method,thevarajan2020breadth}, with SARS‑CoV‑2‑specific T-cells detectable within one to two weeks of infection~\cite{bertoletti2021t}. Our three sampling time points, with median times of 12, 19, and 63 days post symptom onset  (Fig. ~\ref{fig:tcrcovid_fig1}A), correspond to approximately 19, 26, and 70 days post infection (Fig.~\ref{fig:tcrcovid_fig1}A), given that symptom onset for the original Wuhan variant of SARS-CoV-2 lags infection by about 6.5 days~\cite{wu2022incubation,manica2023estimation}. Hence, the first sample time point  lies just after the canonical T-cell expansion apex, whereas later samples probe the chronic phase in which PASC could develop.

To track clonal dynamics within each repertoire, we compared  clones sampled before 20 days  with those after  60 days since symptom onset.  To characterize the different dynamical modes of response, we computed principal components of the  normalized clonal frequency time trajectories for the top rank-1000 clones in each donor. Applying hierarchical clustering on these principal components reveals highly diverse dynamical modes across donors in our large cohort, with some individuals harboring clones with sharp expansion and contraction, and some others lacking any stable high‑frequency clones  (Figs.~\ref{fig:clonal_expansion}A,B;~\ref{fig:SI_tcrcovid_pasc_dynamic_modes},~\ref{fig:SI_tcrcovid_nonpasc_dynamic_modes}). This is in contrast to the similar analysis done in ref.~\cite{minervina2021longitudinal}, in which three dynamical modes were observed to be  largely consistent among the two COVID-19 patients.  

The variability across  individuals in our cohort highlights the importance of individual‑level analyses to quantify T-cell responses. To do so, we used NoisET, a Bayesian method for identifying significantly dynamic clones from longitudinal repertoire samples~\cite{koraichi2022noiset}; see Methods for details.  NoisET compares the extent of changes in clonal frequencies between  repertoires of a given individual sampled at different time points to the variation across replicates within a time point (Method). However, our data lacks biological replicates, and therefore, to make such a comparison, we constructed {\em in-silico} replicate data with Poisson bootstrapping (Methods).  Using NoisET, we were then able to quantify the overall change in the clonal frequency distributions over time, and computed a p-value associated with expansion or contraction  of each clone; clones are considered significantly dynamic if their BH-FDR adjusted $\text{p-value} < 0.01$, with BH-FDR corrections applied separately for each individual (Methods).

Characterizing clonal dynamics between early ($ < 20$ days  post symptom onset) and late time points ($ > 60$ days  after symptom onset), we detected 2,194 expanding and 1,355 contracting TCR$\beta$ clones in the  PASC{$^+$} individuals and 1,015 expanding and 1,045 contracting clones in PASC{$^-$} individuals  (Fig.~\ref{fig:SI_dynamics_volcanos_noiset}). To assess SARS‑CoV‑2 specificity, we compared these clones against a reference set of 138,000 SARS-CoV-2-specific TCR$\beta$ sequences identified using a Multiplex Identification of Antigen-Specific (MIRA) assay~\cite{Nolan2025-fu} (Methods), designating a clone as a match if it fell within a TCRdist of 12 from at least two MIRA entries (corresponding to $\leq$1 amino acid difference in the CDR3 or $\leq$3 differences in the V gene); we chose such a strict  threshold due to a large false positive rate in the MIRA database~\cite{ Nolan2025-fu}. By this measure, 9.4\% of expanding and 15\% of contracting clones in the PASC{$^+$} patients, and 8.3\% of expanding and 17.2\% of contracting clones in PASC{$^-$} patients matched MIRA (Fig.~\ref{fig:clonal_expansion}C), with no significant differences between the two groups (one-tailed Fisher exact test, Bonferroni-adjusted $\text{p-value}>0.05$). As control,  the size‑matched samples of the 200 most abundant clones overlapped MIRA at 9\%, and the random repertoire samples overlapped at about 5\%. These indicate significant enrichment of SARS‑CoV‑2‑specific TCRs in contracting clones against both controls and in expanding clones against only the random repertoire subsamples; see Fig.~\ref{fig:clonal_expansion}C and Table~\ref{table:tcrcovid_mira_dynamics}. Together, these findings suggest that the clones contracting from day 20 to day 60 are preferentially enriched in SARS-CoV-2 reactivity when compared to both controls, while the expanding clones likely represent large, pre‑existing clones whose relative abundance dips during acute infection and rebounds thereafter. \\

\noindent {\bf Differential sequence characteristics  of dynamic clones in PASC{$^+$} and PASC{$^-$}  patients.} Using TCR-OT to identify the differential sequence statistics of the dynamic TCR repertoires  between the PASC{$^+$} and PASC{$^-$}  patients,  we identified 225  expanding and 210 contracting TCRs enriched in  PASC{$^+$} repertoires over those of PASC{$^-$}, and 137 expanding  and 205 contracting TCRs that were enriched in PASC{$^-$} repertoires over PASC{$^+$} repertoires (Fig.~\ref{fig:SI_dynamics_volcanos_ot}). We then identified sequence clusters for the PASC$^+$- or PASC$^-$-enriched expanded  and contracted clones (Figs.~\ref{fig:clonal_expansion_OT}A,~\ref{fig:SI_tcrcovid_tcrot_expanding_clusterfit},~\ref{fig:SI_tcrcovid_tcrot_contracting_clusterfit}), with clusters varying widely in size and the number of individuals represented in them (Figs.~\ref{fig:clonal_expansion_OT}B,C).

Cross-referencing  sequences of these differentially enriched clusters with VDJdb~\cite{shugay2018vdjdb,goncharov2022vdjdb}  showed a significant enrichment of Influenza A-specific TCRs in the contracting PASC{$^+$} repertoires compared to the  PASC{$^-$} repertoires (Mann-Whitney $U = 170$, $n_+ = 13$, $n_- = 16$, Bonferroni-corrected $\text{p-value} = 0.016$ (Figs.~\ref{fig:clonal_expansion_OT}D,E). These contracting TCRs were present in 15 of PASC{$^+$} individuals. Notably, PASC{$^+$} individuals whose contracting TCRs match Influenza A-specific TCRs (Fig.~\ref{fig:clonal_expansion_OT}E) exhibit significantly greater WHO ordinal scale at early times ($t<20$ days)  than PASC{$^-$} individuals in the same category---a difference larger than that observed between the broader PASC{$^+$} and PASC{$^-$} repertoires (p-value $= 0.015$, by comparing Mann-Whitney test statistics against that of the bootstrapped paired sample sets drawn from the general PASC{$^+$} and PASC{$^-$} repertoires); see Table~\ref{table:SI_VDJDB_WHO} for details. As these samples were collected in early 2020, the most plausible explanation is a recent or concurrent influenza infection, since there is little evidence supporting bystander or viremia activation associated with influenza. This finding suggests that the elevated severity of illness associated with viral co-infections~\cite{Alosaimi2021-mc,Achdout2021-pt} could influence PASC pathogenesis and warrants systematic investigation.

Lastly, we characterized the sequence motifs associated with each of the identified TCR-OT clusters. Sequence logos constructed from neighborhoods within a 48 TCRdist radius around the focal clones of significant TCR-OT clusters associated with contracting clones revealed recurrent TRBV gene usage--specifically TRBV9 and TRBV6-6 in  PASC{$^+$} and TRBV7-9 and TRBV6-1 in PASC{$^-$}  patients (Figs.~\ref{fig:clonal_expansion_OT}F,~\ref{fig:SI_tcrcovid_tcrot_dynamics_motifs}). These patterns mirror TRBV gene preferences previously linked to SARS-CoV-2 reactivity~\cite{pogorelyy2022resolving}.

\begin{figure*}[htbp]
\includegraphics[width=0.95\textwidth]{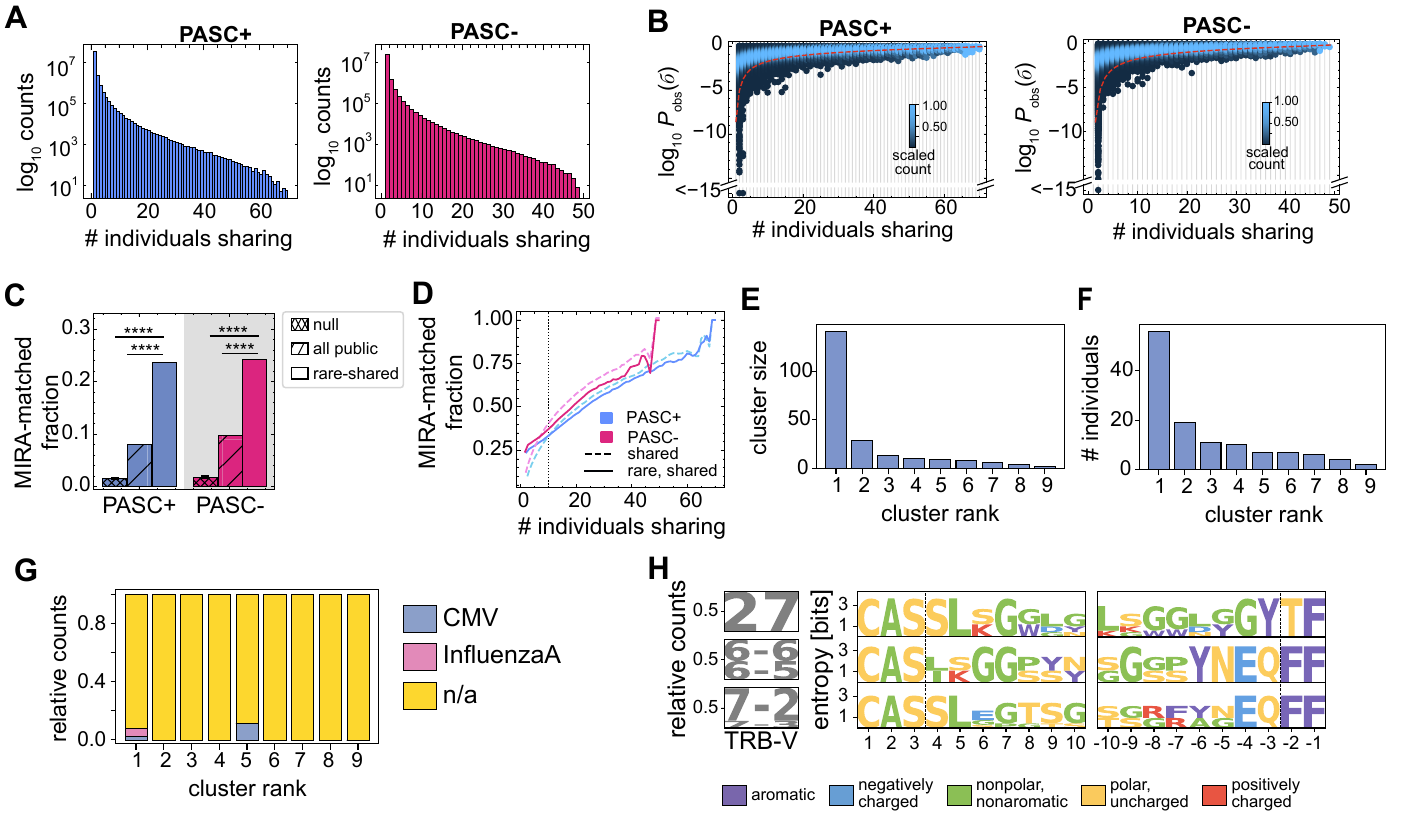}
\caption[Sharing of TCR repertoires.]{{\bf Sharing of TCR repertoires.}
 {\bf(A)} The sharing spectrums of the PASC{$^+$} (left) and PASC{$^-$} (right) patients are shown as histograms and illustrate the number of TCRs  which are incident over a given number of individuals. 
  {\bf(B)} The density plot shows the distribution of $\log_{10} \pobs$  for TCRs shared in a given number of individuals  for the PASC{$^+$} (left) and  PASC{$^-$} (right) groups.  The scaling of the counts is set for each number of individuals (column) separately, with the maximum of the density given a value of one.
  The parametric  curve (the red, dashed line) indicates an equi-$\pshare$ line (sharing probability), such that $0.5\%$ of the repertoire lies under it;  sequences below this curve are assumed to be outlier public clones and rare  (Methods).  {\bf(C)} The bar plots report the mean fraction of MIRA matches for the rare public TCR subsets (solid, right bars), the fraction of matches for all public TCR subsets without restricting to rare receptors (dashed, middle bars), and fraction of matches for 200 realizations of repertoire subsets drawn randomly from the bulk repertoires (null) consistent with the $\pobs$ distribution of the rare subsets (hatched; left bar), shown separately for PASC{$^+$} (blue) and  PASC{$^-$} (magenta) groups.  The error bar on the hatched bars indicates one standard deviation of variation.
  ****:~Bonferroni-corrected $p$-value $< 10^{-4}$.   {\bf(D)}  The fraction of MIRA matches for TCRs shared among the minimum number of individuals is shown for PASC{$^+$} (blue) and  PASC{$^-$} (magenta) groups.  
  The dashed lines show the expectation from the entire shared repertoire whereas the solid line shows the incidences for the  subset of rare-shared TCRs (those below the red line in (B)). The vertical dotted line indicates  where the two curves first approximately intersect (at around 10 individuals).
  {\bf (E-H)} TCR-OT analysis: We applied TCR‑OT at the level of amino acid sequences  to identify differential motifs between rare-shared receptors in PASC{$^+$} and PASC{$^-$} groups. Among rare public clonotypes, we found 9 clusters enriched in PASC{$^+$} over  PASC{$^-$} repertoires (and none the other way round) that contain $>20\%$  enriched members. 
 {\bf (E)} The amounts of independent recombinations in these PASC{$^+$} enriched TCR-OT  clusters and {\bf (F)} the number of individuals represented in a cluster  are shown. Cluster rank corresponds to the size of the clustering.  {\bf (G)} The stacked bar graph show the distribution of epitopes of  TCRs in VDJdb that were within 24 TCRdist of a constituent TCR in a TCR-OT cluster  and whose donor HLA group matched to at least one HLA group of the individual from which the receptor originated.   {\bf(H)} Sequence logos describing the composition of amino acid sequences are shown for the significant TCR-OT clusters obtained from these rare public  TCRs in PASC{$^+$}-enriched TCR-OT clustered.  Logos were constructed from focal sequence(s) and members in the cluster that were within 48 TCRdist of the focal sequence(s), and logos are shown only if 10 sequences were present in the constrained neighborhood for a cluster.
\label{fig:tcrcovid_sharing}
}
\end{figure*}

\section{Impact of PASC on the composition of the  TCR public repertoires}

\noindent {\bf Identifying rare public clones.}  Despite the vast diversity of TCRs within repertoires, we observed a substantial number of identical public TCR clones shared among individuals in both PASC{$^+$} and PASC{$^-$}  groups (Fig.~\ref{fig:tcrcovid_sharing}A). Although sharing of TCRs can signal convergent responses to a common immune challenge, it can also arise due to convergence recombination leading to the production of the same receptor amino acid sequence. Therefore, it is imperative to use robust statistical models that can reveal the significance of  sharing for different TCRs beyond the chances of convergent recombination. To do so, we employed the softwares soNNia (version 0.3.2)~\cite{isacchini2021deep} and IGoR (version 1.4) \cite{marcou2018high}) to  train generation and  selection models for each individual's repertoire to characterize the probability to observe a TCR sequence $\sigma$ in the periphery $P_\text{post}(\sigma)$.

Intuitively, sharing is more likely among commonly occurring receptors (i.e., with high $P_\text{post}(\sigma)$) and within groups with larger sampling (Methods). Therefore, rare receptors (i.e., with a low $P_\text{post}(\sigma)$) that are shared among many individuals with a common immune challenge can signal commonality in function, as observed previously for both T-cell and B-cell repertoires~\cite{elhanati2018predicting,pogorelyy2018method,montague2021dynamics,ruiz2023modeling}. Therefore,  probability that a TCR sequence $\sigma$ is shared $\pshare(m; M, \sigma)$ depends on  the number individuals who share that receptor $m$,  the TCR's occurrence probability $P_\text{post}(\sigma)$, and the total of individuals $M$ in a group; the exact expression for this sharing probability is derived in the Methods. We define rare-shared clones as those with $\pshare(m; M, \sigma)<0.5\%$ (Fig.~\ref{fig:tcrcovid_sharing}B).

Using this procedure yielded 19,847 rare public TCRs from the 3.97 million shared clones in the PASC{$^+$} repertoires, and 13,417 from the 2.68 million in the PASC{$^-$}'s (Fig.~\ref{fig:tcrcovid_sharing}B). Roughly one quarter of these (23.7\% in PASC{$^+$} and 24.3\% in PASC{$^-$}) matched SARS-CoV-2-specific sequences in the MIRA database~\cite{ Nolan2025-fu} (Fig.~\ref{fig:tcrcovid_sharing}C).  The two groups shared 5,762 of these rare public clones (29\% of the PASC{$^+$} subset and 42\% of the PASC{$^-$}). Excluding this overlap, the rare public clones in PASC{$^+$} individuals showed a significantly higher proportion of MIRA matches than PASC{$^-$} (18\% vs. 14.4\%; one-sided Fisher exact test, Bonferroni-corrected p-value $= 4.18 \times 10^{-12}$), suggesting stimulation of a potentially more diverse repertoire of SARS-CoV-2-specific  T-cells in  PASC{$^+$} patients.  

As a control, we repeated our analysis for all the public clones, without restricting to the rare receptors. We found that only 8.1\% (in PASC{$^+$})  and 9.7\% (in PASC{$^-$}) of the  public clones matched the MIRA, indicating that for most public clones, sharing is largely driven by   convergent recombination rather than responses to common antigens (Fig.~\ref{fig:tcrcovid_sharing}C). In addition, we also compared the fraction of MIRA matches in the rare-shared TCRs with the  fraction of  MIRA matches from randomly drawn subsets sampled such that the probability of observing them in the repertoires $\pobs$  is similar to that of the rare-shared subset. We observed less than 2\% of this randomized set ($1.55\% \pm 0.09\%$ in PASC{$^+$}  and $1.81\% \pm 0.11\%$ in PASC{$^-$}) matched  MIRA (Fig.~\ref{fig:tcrcovid_sharing}C).

Lastly, as we required TCRs to be shared across progressively more individuals, the MIRA enrichment gap between all public and rare public receptors narrowed and disappeared once a TCR was present in more than ten individuals (Fig.~\ref{fig:tcrcovid_sharing}D).  This suggests that high-publicness alone becomes a sufficient indicator of response to a common antigen, as convergent recombination becomes very unlikely at that scale. Taken together, our analysis indicates that rare public TCRs are significantly enriched for SARS‑CoV‑2 specificity beyond repertoire‑wide expectations ($Z$-test, Bonferroni-corrected $p = 0$ (below machine precision)), but that PASC status does not significantly impact the specificity of rare public TCR to SARS-CoV-2.\\

\noindent {\bf Differential sequence characteristics of rare public clones in PASC{$^+$} and PASC{$^-$}  groups.} Having identified the rare public TCRs, we next used the TCR-OT algorithm to distinguish sequences unique to each group. Because rare public TCRs are defined by enrichment within a group, applying TCR-OT at the nucleotide level (as we have done previously) would trivially flag most sequences as group-enriched; we therefore applied it at the amino acid level to detect differing motif  within the rare public subsets.

We found 51 (0.26\%) and 20 (0.15\%) sequences in the PASC{$^+$} and PASC{$^-$} rare, shared subsets, respectively, to be significantly enriched in their respective group (TCR-OT randomization test, BH-FDR adjusted p-value$ < 0.05$), with only 9 PASC{$^+$} and 0 PASC{$^-$} TCR-OT clusters had at least 20\% of their constituents called significant. Motifs constructed from PASC{$^+$}-enriched clusters show distinct sequence patterns  (Fig.~\ref{fig:tcrcovid_sharing}H), and these clusters vary widely in size and  individual representation (Figs.~\ref{fig:tcrcovid_sharing}E,F). Matching these clusters to VDJdb revealed slight specificity toward CMV and Influenza A in a few cases, though most sequences did not match any VDJdb entry (Fig.~\ref{fig:tcrcovid_sharing}G).

\section{Discussion}

Post-acute sequelae of SARS-CoV-2 infection (PASC) is associated with persistent immune dysregulation, yet the specific immunological features that distinguish individuals with PASC from those who recover fully remain poorly defined~\cite{Davis2023-qe,Al-Aly2024-vc, Yin2024-uw}. Here, we profiled longitudinal bulk TCR$\beta$ repertoires from 120 individuals in the INCOV cohort~\cite{su2020multi}  and applied robust statistical frameworks to identify candidate T-cell receptors associated with PASC status. 

Global repertoire statistics, i.e., V and J gene usage, CDR3 length distributions, and clonality, did not differ significantly between PASC{$^+$} and PASC{$^-$} groups. This contrasts with chronic persistent infections such as CMV, where immunodomiNaNt expansions can occupy a large fraction of the circulating T-cell pool and produce  detectable changes in repertoire composition~\cite{Weekes1999-rx,emerson2017immunosequencing}. However, 
locally enriched sequence motifs identified via TCR-OT \cite{olson2022comparing} revealed differential sequence signatures between the two groups~\cite{isacchini2024local}.

Longitudinal analysis of clonal dynamics revealed that clones contracting after day 20 post-symptom onset were significantly enriched for SARS-CoV-2 specificity in both groups relative to controls, consistent with canonical contraction of acutely expanded virus-specific effector clones.  No significant difference in this enrichment was observed between PASC{$^+$} and PASC{$^-$} individuals, suggesting that the magnitude of the acute antigen-specific response does not stratify by long-term clinical outcome. However, Influenza A-specific TCRs were significantly enriched among the contracting clones of PASC{$^+$} individuals relative to PASC{$^-$}, concentrated in participants with higher early disease severity. Given the early-pandemic timing of sample collection, this likely reflects concurrent influenza co-infection, and is consistent with reports that viral co-infections  elevate COVID-19 severity~\cite{Alosaimi2021-mc,Achdout2021-pt}, which may influence PASC pathogenesis.

Analysis of the public TCR repertoire showed that rare public clones, i.e., those shared beyond chance expectations given receptor generation biases, were enriched for SARS-CoV-2 specificity, far above the rate observed for all public clones, or randomized controls, confirming convergent recombination due to a common immune challenge. After excluding clones shared between groups, PASC{$^+$} individuals harbored a modestly but significantly higher fraction of SARS-CoV-2-matched rare public clones than PASC{$^-$} individuals, suggesting a more persistent or broader antigen stimulation in PASC{$^+$}. 

Several limitations for this study should be noted. The data was sampled  during the first wave of the pandemic with the ancestral SARS-CoV-2 strain, prior to widespread vaccination, and it is unclear whether our conclusions  extend to infection with subsequent viral variants or to the post-vaccination immune landscape, in which the baseline of SARS-CoV-2-specific memory is substantially different. Only the TRB chain was sequenced; paired $\alpha\beta$ data would considerably increase the precision of epitope-level inference~\cite{Pogorelyy2026-hs}. The study was confined to peripheral blood, whereas tissue-resident T-cell populations at sites of PASC pathology may harbor more informative signals~\cite{Littlefield2022-bn, Santa-Cruz2023-da}. Finally, PASC itself is phenotypically heterogeneous with evolving definitions, and the group size, while on the larger end for bulk repertoire studies, limits statistical power for subgroup analyses stratified by symptom cluster or severity. Resolving the functional roles of the candidate TCRs identified here--whether they contribute to PASC pathogenesis, participate in its resolution, or serve as biomarkers of divergent immune trajectories--will require larger prospective cohorts, tissue sampling, paired chain sequencing, and integration with single-cell transcriptomic and functional readouts.

\section{Acknowledgements}
This work has been supported by the  National Institutes of Health award R03AI175977, and in part by the CAREER award from the National Science Foundation grant 2045054 (A.N., Z.M., A.T., R.M.G.), and the National Institutes of Health MIRA award R35~GM142795 (A.N., Z.M., A.T.). Additional support for A.T. was provided by the Mahan Fellowship from the Fred Hutchinson Cancer Center. This work is also supported, in part, through the Departments of Physics and Applied Mathematics and the College of Arts and Sciences at the University of Washington (UW). The numerical analyses in this work  were completed on Hyak, the UW's high performance computing cluster, which is  funded by the UW student technology fee.  R.M.G. acknowledges the REU program at the Department of Physics at the University of Washington, which is supported by the National Science Foundation grant 2243362.

\newpage{}
\clearpage{}
\onecolumngrid


\noindent {\bf\Large Materials and Methods} 

{\subsection*{Data and code availability}
Part of the TCR repertoire data was published in ImmuneCODE database~\cite{Nolan2025-fu}, generated as part of the study from ref.~\cite{su2022multiple}. The complete data along with meta data on patient's status are published along this manuscript and  can be accessed through: \href{https://doi.org/10.5281/zenodo.19836291}{https://doi.org/10.5281/zenodo.19836291}. All codes for data processing and statistical analysis can be found at: \href{https://github.com/StatPhysBio/PASCCC}{https://github.com/StatPhysBio/pasc}.
}

\subsection*{T-cell receptor variable beta chain sequencing}
High-throughput TCR$\beta$ sequencing were performed as previously reported~\cite{Robins2009-wr,Carlson2013-jr,Chapuis2019-uu, su2022multiple,Nolan2025-fu}. Briefly, DNA was extracted from T cells and TCR$\beta$ CDR3 (complementarity determining region 3) regions were sequenced using the immunoSEQ\textregistered Assay (Adaptive Biotechnologies, Seattle, WA), a multiplex PCR-based method that amplifies and characterizes CDR3 rearranged sequences, with a built-in rigorous PCR amplification bias control and quality assurance. 

\subsection*{Patient symptoms }
We quantified symptom differences using the Jaccard distance with Ward clustering, identifying five coherent clusters: gastrointestinal (diarrhea, nausea), anosmia/dysgeusia (loss of taste, smell), abdominal pain, respiratory (cough, sputum), and neurological (fatigue, shortness of breath, exercise intolerance). To test associations between sex or age and symptom presentation, we used mutual information, which measures statistical dependence between variables:
\begin{equation}
	I(x,y) = \sum_{x,y} P(x,y) \log_2 \frac{P(x,y)}{P(x)P(y)}
\end{equation}
Significance was assessed via randomization tests with 1,000 permutations per comparison, accounting for finite-size bias and multiple testing correction. No significant associations were found between symptoms and either sex or age (Fig.~\ref{fig:tcrcovid_fig1}B).

\subsection*{Clustering and grouping of HLA alleles}
HLA alleles were grouped based on peptide-binding preferences in accordance with established hierarchical clustering methods~\cite{Pierini2018-py}. Specifically, for each of the six allele categories (A, B, C, DRB1, DP, DQ), we performed k-means clustering over each allele's peptide binding cleft sequences, using a Grantham distance~\cite{Pierini2018-py} between amino-acid residues. This results in 6 groups for each HLA allele category, as detailed in Dataset~1. For all analyses in the manuscript, HLA alleles are  grouped according to their peptide binding cleft  similarity.

\subsection*{Clonality of repertoires}
Clonality quantifies the imbalance of clone distributions in a sample based on clonal abundances. We defined clonality using the normalized Shannon entropy,
\begin{equation}
    \mathcal{C} = 1 + \frac{\sum_{i = 1}^N f_i \log f_i}{\log N},
\end{equation}
where $N$ is the number of clones in a sample and $f_i$ is the frequency of clone $i$ in the repertoire. Normalization ensures comparability across samples of different sizes. $\mathcal{C} = 0$ indicates each clone has the same abundance whereas $\mathcal{C} \approx 1$ indicates domiNaNce by a few clones. Other definitions of clonality have been used to understand lymphomas~\cite{wu2012high}, cytomegalovirus (CMV) and graph-versus-host disease~\cite{calderin2023cmv}, and how HLA shapes the TCR repertoire~\cite{krishna2020genetic}.

Repertoire clonalities of PASC$^{+}$ and PASC$^{-}$ individuals, stratified by time since symptom onset and by sex, showed no significant differences, nor did fold changes in clonality across time points (Mann–Whitney U test with Bonferroni correction, $p > 0.05$; Fig.~\ref{fig:SI:tcrcovid_composition}E). This suggests that PASC does not modulate the distribution of clone sizes in a manner distinct from acute COVID-19.

\subsection*{Comparing TCR feature statistics across cohorts}
We compared  repertoire statistics for  PASC{$^+$} and PASC{$^-$} groups to the 379 repertoires of CMV{$^-$} healthy individuals from Emerson cohort \cite{emerson2017immunosequencing}. We aggregated gene usage and CDR3 amino acid length data from the INCOV cohort (differentiated by PASC status) and the Emerson cohort to look for significant differences in gene usage. In the Emerson cohort, we grouped alleles of the same gene together; the INCOV cohort does not contain allele-level data. We performed hypothesis testing using Welch's t-test, and adjusted p-values using Bonferroni correction, multiplying each p-value by the number of tests performed (24 for J gene usage and 72 for V gene usage). Three V genes represented in the INCOV cohort were not present in the Emerson data. The comparisons are shown in Fig.~\ref{fig:SI:tcrcovid_composition}A-D, and the results of the hypothesis testing are reported in Tables~\ref{table:vgene_ttests}--\ref{table: cdr3_ttests}.

\subsection*{Identifying differential sequence motifs between repertoires with TCR-OT}
To characterize the differential features of repertoires between various subsets of  PASC{$^+$} and PASC{$^-$} groups, we used a nonparameteric, interpretable statistical method based on optimal transport (TCR-OT) \cite{olson2022comparing} that balances interpretability with biologically meaningful predictions.
TCR-OT takes as input two repertoire datasets---$\mathcal{R}_1$, $\mathcal{R}_2$---each containing deduplicated, independent V(D)J recombination events, i.e., TCR sequences defined by their  V gene, J gene, the CDR3 nucleotide sequence, and the individual from which they were sampled.
Both datasets undergo preprocessing which deduplicates the TCRs at the level of amino acid sequences (V gene and CDR$\beta$  amino acid sequence) and retains their multiplicities.
Probability mass distributions for each dataset, $\mathcal{P}_1, \mathcal{P}_2$, are formed by normalizing the multiplicities of the amino acid sequences in their respective datasets.
A distance matrix $D$ of TCRdists~\cite{dash2017quantifiable,mayer2021tcr}  is computed between all pairs of TCRs in $\mathcal{R}_1$ and $\mathcal{R}_2$.
Using the probability mass distributions and the distance matrix $D$, TCR-OT finds the optimal transport plan $\Pi^\star$ that minimizes the transport cost (distance) for mapping one repertoire onto the other, while keeping the marginal distributions consistent with that of the two repertoire data.  The Wasserstein distance is then computed as the corresponding  distance between the two repertoires under the optimal transport plan, as the  Hadamard product of $\Pi^\star$ and $D$,
\begin{equation}
    W = \Pi^\star \odot D
\end{equation}
and represent the minimum ``effort" that must be exerted to move one repertoire to another.

For a given TCR sequence $\sigma$ in the repertoire $\mathcal{R}_1$, its set of neighbors is given by 
\begin{equation}
    \mathcal{N}(\sigma) = \{\sigma' \, | \, \tcrdist(\sigma, \sigma') \leq t\}.
\end{equation}
We mirror \cite{olson2022comparing} and let $t = 48$, which amounts to a difference of about 4 amino acids in the CDR3 sequences or 12 amino acid differences between the V genes of $\sigma$ and $\sigma'$.
The TCR-OT enrichment statistic $s(\sigma)$ for TCR $\sigma$ is computed as the overall Wasserstein distance between a neighborhood of TCRs around the TCR, $\mathcal{N}(\sigma)$, and all the TCRs in $\mathcal{R}_2$
\begin{equation}
    s(\sigma) = N_1 \sum_{\sigma' \in \mathcal{N}(\sigma)} \delta(\sigma')_i \mathbf{1}_j W_{ij},
\end{equation}
where $\delta(\sigma')$ is a vector of 0's except at the position indexed by $\sigma'$ where it is 1, and scaling the statistic by the size of the repertoire $N_1$ ensures the statistic is comparable across datasets of different sizes. Enrichment statistics for sequences in $\mathcal{R}_2$ are computed similarly.

To quantify the significance of a TCR's enrichment, TCR-OT performs a randomization test.
Each realization of the randomization test shuffles the labels that specify whether an independent recombination comes from $\mathcal{R}_1$ or $\mathcal{R}_2$, and then recomputed the  enrichment scores  on this shuffled dataset. After generating many realizations, a distribution of enrichment scores is obtained for each input TCR, characterizing the enrichment scores we would expect if $\mathcal{R}_1$ and $\mathcal{R}_2$ were samples from an identical population of TCRs. A $p$-value is calculated for each TCR sequence using the quantile of its observed enrichment score---the score computed using the unshuffled labels---relative to this distribution. To identify the significantly enriched TCRs, we perform randomizations until each TCR was sampled in their respective dataset 500 times, and assign those TCRs with $p$-value $< 0.05$ (after BH-corrections~\cite{hochberg1990more}) as significantly enriched.

TCR-OT also constructs clusters of TCRs that appear to be enriched in their respective dataset.
Its algorithm is as follows.
Consider $\mathcal{R}_1$ which can be partitioned into a set which contains TCRs which have been clustered already $\mathcal{R}_{1,c}$ and those which haven't $\mathcal{R}_{1,\tilde{c}}$.
Initially $\mathcal{R}_{1,c} = \emptyset$ (is empty) and $\mathcal{R}_{1,\tilde{c}} = \mathcal{R}_{1}$.
Consider the most enriched TCR in $\mathcal{R}_{1,\tilde{c}}$: $\sigma^\star$.
Moving away from $\sigma^\star$ in steps of $\Delta r$ TCRdists until $r_{\max}$, we identify the subset $\mathcal{A}_i $ of TCRs within the $r_i$ and $r_i + \Delta r$ annulus and compute the mean enrichment $m_i$
\begin{equation}
	m_i(\sigma^\star) = \langle s(\sigma) \rangle_{\mathcal{A}_i(\sigma^\star)},
\end{equation}
where $\langle\cdot\rangle_{\mathcal{A}_i(\sigma^\star)}$ indicates expectation of the TCRs in the set $\mathcal{A}_i $.
Using segmented linear regression \cite{muggeo2003estimating}, a breakpoint in the mean enrichment vs. TCRdist annulus radii relationship can be detected which demarcates when the mean enrichment has saturated; these curves are shown in Figs.~\ref{fig:tcrcovid_tcrot_top50_clusterfit},~\ref{fig:SI_tcrcovid_tcrot_expanding_clusterfit},~\ref{fig:SI_tcrcovid_tcrot_contracting_clusterfit}.
All TCRs with TCRdist less than this breakpoint are clustered with the focal TCR $\sigma^\star$, and $\mathcal{R}_{1,c}$ and $\mathcal{R}_{1,\tilde{c}}$ are updated accordingly.
TCRs clustered together in this manner are believed to signal similar function since focal sequences were characterized as atypical with respect to the reference repertoire via the TCR-OT enrichment statistic.
TCR-OT's clustering proceeds by finding a pre-specified number of clusters or until the segmented linear regression fails to identify a breakpoint.
In our analysis, $\Delta r = 5$ TCRdist with maximum radius $r_{\max} = 200$ were chosen.
We retained only clusters with cluster radius $r_{\mathrm{cluster}} \leq 120$, corresponding to a difference of at most 10 amino acids in the CDR3 sequence assuming the same V gene, for downstream analyses.
This choice balances sequence diversity within a cluster with intolerance toward constituent cluster sequences being too functionally dissimilar to the focal sequence.
Moreover, segmented linear regressions producing $r_{\mathrm{cluster}} > 120$ ensued from performing inference on sparse subsets of the dataset, so this conservative choice ensures we examined only robust clusters downstream.
Most cluster radii inferred in this study were between 50 and 100 TCRdist (Figs.~\ref{fig:tcrcovid_tcrot_top50_clusterfit},~\ref{fig:SI_tcrcovid_tcrot_expanding_clusterfit},~\ref{fig:SI_tcrcovid_tcrot_contracting_clusterfit}), which
 corresponds to roughly at most 4 to 8 amino acid differences, respectively, in the CDR3 sequences relative to the cluster focal sequence.

\subsection{Statistical model for clonal dynamics with NoisET}

 Bulk TCR sequencing data provides a noisy representation of the underlying repertoire in an individual due to undersampling. Therefore, quantifying the dynamics of TCRs over time (e.g. in response to an immune challenge) requires a statistical model to separate variation due to sampling from variation due to underlying population change. To address this challenge, we employed NoisET, a Bayesian method for detecting differential expansion and contraction, to identify significantly dynamic clones from the entire functional repertoires of the aforementioned sets of individuals~\cite{koraichi2022noiset}. 
 
To characterize variation due to sampling, NoisET uses biological replicates of repertoires to create a null model representing typical variation in read counts due to sampling and potential sources of experimental error. It  then detects significantly contracted or expanded clones as those whose fold changes in abundance from the earlier to the later time point fluctuated beyond that expected by the null models. Clone abundances are modeled as a power-law distribution, 
\begin{equation}
  \rho(f) = Cf^\alpha
\end{equation}
where $f$ is the frequency and $C$ and $\alpha$ are constants.
NoisET allows the user to choose between three forms of noise model: a Poisson model, which infers the exponent of the power law distribution $\alpha$ and the true minimum clonal frequency in the repertoire, $f_{min}$; a negative binomial model, which infers $\alpha$, $f_{min}$, and the parameters of the negative binomial distribution; and a mixed negative binomial/Poisson model, which infers the same parameters as the negative binomial distribution,  plus an additional parameter $M$ which accounts for overdispersion. 

After calculating a null model for each time point, yielding two sets of null model parameters $\Theta_1$ and $\Theta_2$, NoisET then calculates the probability of observing a clone with abundance $n(t_1)$ and $n(t_2)$ at times $t_1$ and $t_2$, respectively, using the null model parameters. NoisET also infers the fraction of responding (dynamical) clones $\gamma$ and the characteristic log-change of such clones in the repertoire  $\bar \eta$,  from the overall change in the clonal frequency distributions over time.

These inferred parameters can then be used to test for expansion by calculating a one-sided $p$-value using the posterior distribution; $\eta$ represents the log-fold change of the clone frequency between time 1 and time 2.
\begin{equation}
	p-\text{value} = P(\eta \leq 0 | n(t_1), n(t_2), \gamma, \bar{\eta}, \Theta(t_1), \Theta(t_2)).
\end{equation}
Consistent with this definition, contraction is tested by switching the counts and null model parameters by sending $t_1 \rightarrow t_2$ and $t_2 \rightarrow t_1$. 

In the absence of biological replicates in this study, we opted to generate {\it in silico} replicates via Poisson bootstrapping. Specifically, given a repertoire sampled at a single time point, we produced two replicates by Poisson resampling using the observed TCR's read count as the Poisson rate. For each sampled repertoire, we generated 100 sets of {\it in silico} paired replicates, learned 100 NoisET Poisson noise models, and took the mean of the resulting parameters for that sample as input for NoisET to model the sampling noise in the data. For the null model, we used the Poisson option, as it was the most parsimonious and produced the most consistent results across \textit{in silico} replicates. 

We considered clones significantly dynamic if their BH-FDR adjusted $\text{p-value} < 0.01$, with BH-FDR corrections applied separately for each individual.

\subsection*{Training generation and selection models}
We used IGoR (version 1.4) \cite{marcou2018high} to learn a model of VDJ recombination for each individual by pooling all their observed unproductive sequences.
Since the provided annotations did not contain a CDR3 nucleotide sequence for unproductive sequences, IGoR was used to align the full TCR nucleotide sequence against genomic templates to identify the nucleotide CDR3 sequence.
After this alignment, the unproductive sequences were deduplicated at the level of  CDR3 nucleotide sequence, and V and J alleles to retain only those sequences which most likely came from independent recombination events.
IGoR models were initialized from the default {\trb} IGoR model and trained using 10 epochs of expectation-maximization. The trained model can be used to quantify the  probability for generating a TCR $\pgen(\sigma)$, parametrized by   V,  D, and  J gene choices as well as the insertion and deletion profiles at the VD and DJ junctions, and to generate new sequences, consistent with  $\pgen$.

Next, we used soNNia (version 0.3.2)~\cite{isacchini2021deep} to model the effects of functional selection (be it from thymic selection or response in the memory repertoire).
soNNia infers selection factors $Q^\theta(\sigma)$ to estimate the fold change difference between the probability $\ppost(\sigma)$ to observe a productive sequence in the periphery and the probability $\pgen(\sigma)$ of generating the sequence,
\begin{equation}
	\ppost(\sigma) = \frac{1}{Z} \pgen(\sigma) Q^\theta(\sigma).
\end{equation}
$Z$ is a normalization factor, and $Q^\theta(\sigma)$ is a neural network which takes as input  V gene and  J gene usages,   CDR3 sequence length, and   CDR3 amino acid composition~\cite{isacchini2021deep}.

The soNNia models were trained for each individual's repertoire using the deep architecture with 150 epochs, 500,000 $\pgen$ sequences, $L_2$ regularization of $3 \times 10^{-4}$, joint V-J features, and a batch size of $5,000$.
Training datasets were constructed from productive TCRs whose   CDR3 amino acid sequences began with a cysteine and whose   CDR3 amino acid sequences lengths did not exceed 30 residues. To limit biases due to expansion, the training datasets were deduplicated at the level of   CDR3 nucleotide sequences,  V gene, and  J gene. We trained a soNNia model for each individual, pooling their repertoires sampled across time, to characterize selection due to individual HLA restrictions as well as different immune responses mounted over the course of their infections. When learning a soNNia model for each sampled repertoire, $\pgen$ sequences were generated from the associated individual's bespoke IGoR model.

\subsection*{Sharing statistics and  identifying public repertoires}
Sharing of sequences can be due to biases in V(D)J recombination, similar HLAs and convergent thymic selection, convergent pathogenic selection, or experimental biases, and using these models allows us to quantify our expectations of a sequence shared due to V(D)J recombination and the average effects of selection~\cite{elhanati2018predicting,pogorelyy2018method,montague2021dynamics,ruiz2023modeling}.

We studied the sharing of TCRs among individuals by aggregating repertoires over time for each individual and deduplicating the data at the level of V gene, J gene,  CDR3 amino acid sequence, and individual. Let $\sigma$ be a TCR clone defined here  as the V gene, J gene, and CDR3 amino acid sequence.
The observation of $\sigma$ in each individual is a Bernoulli trial.
Because each $\sigma$ has a different probability of being observed in individual, the number of times a sequence is shared among individuals is modeled as a Poisson binomial process.
We approximate the Poisson binomial distribution characterizing these phenomena using a binomial distribution parameterized by the maximum amount of possible incidences and the average probability of observing a sequence~\cite{choi2002approximating}.
This approximation will yield conservative estimates when quantifying outliers because the binomial distribution is more dispersed than the Poisson binomial distribution.
The probability of observing $\sigma$ at least once in individual $i$'s repertoire is
\begin{equation}
	\pobs(\sigma; i) = 1 - (1 - \ppost(\sigma; i))^{N_\nt(i)},
\end{equation}
where $\ppost(\sigma; i)$ is the probability of observing $\sigma$ in the periphery estimated using the soNNia model associated with individual $i$, and $N_\nt(i)$ is the number of unique {\trb} nucleotide sequences, i.e., sequences defined by the V gene, J gene, and  CDR3 nucleotide sequence, in individual $i$'s pooled repertoire.
We next computed the probability of sharing $\sigma$ in $m$ out of $M$ individuals:
\begin{equation}
	\pshare(m; M, \sigma) = {M \choose m} \rho(\sigma)^m (1 - \rho(\sigma))^{M - m}.
\end{equation}
The probability $\rho(\sigma)$ is the average probability of observing $\sigma$ in each individual at least once,
\begin{equation}
	\rho(\sigma) = \frac{1}{M} \sum_{i = 1}^{M} \pobs(\sigma; i).
\end{equation}
To identify outliers in the $\pshare$ distribution, we fix a threshold $c$ such that
\begin{equation}
	c = {M \choose m} q(m)^m (1 - q(m))^{M - m},
\end{equation}
with $q(m) \in [0, 1]$, a probability tuned for each $m$. This gives a parametric curve of probabilities $q(m)$ versus sharing number $2 \leq m \leq M$, and sequences below this curve are assumed to be outliers and rare.
We chose $c$ such that $0.5\%$ of the repertoire lie under the parametric curve $q(m)$.

\subsection*{Creating a control for the subset of rare clones}
We mirror ref.~\cite{ruiz2023modeling} for creating a control cohort and summarize their method here.
We recall that constituents of the repertoire studied for sharing are defined by their  V gene,  J gene, and   CDR3 amino acid sequence.
To create a control relative to the subset of clones identified as rare, we want to sample from the bulk {\trb} repertoire such that we reproduce the $\pobs(\sigma)$ distribution of the rare clones.
We histogram the $\log_{10} \pobs(\sigma)$ distribution of the rare clones evenly with $n$ bins, yielding a height $h_i$ of counts in each bin; here, we use 500 bins spaced evenly between $-40$ and $0$.
We bin the $\log_{10} \ppost(\sigma)$ of the sequences in the deduplicated bulk {\trb} repertoire and sample $h_i$ {\trb} sequences without replacement in each respective bin.
We then check the overlap of those sequences with the MIRA database (Fig.~\ref{fig:tcrcovid_sharing}D).

\subsection*{Graphical analysis with Gephi}
We used Gephi (v.10.1)~\cite{bastian2009gephi} to create undirected network graphs for the TCR-OT clusters (Figs.~\ref{fig:tcrcovid_tcrot_composition},\ref{fig:clonal_expansion_OT}).
For each given TCR-OT cluster, edges connected nodes if nodes were within 48 TCRdist of each other, and all nodes in a cluster had edges to the focal sequences of the cluster (those with the highest enrichment), save for self-edges.
To arrange the clusters, we used the ForceAtlas 2 layout with stronger gravity enabled, gravity set to 1, and a scaling of 10.
Nodes sizes range from 2 to 6, depending on the strength of their BH-corrected $-\log_{10}$ significance of their TCR-OT enrichment statistic.

\subsection*{Matching of TCRs  into VDJdb}
The VDJdb (dated 2024.06.13)~\cite{shugay2018vdjdb,goncharov2022vdjdb} was downloaded from the vdjdb-db GitHub repository.
We restricted the database to TCRs from humans and which were detected as reactive to SARS-CoV-2, EBV, CMV, and InfluenzaA.
Additionally, we removed 10x Genomics data known to be contaminated~\cite{croce2024deep}, i.e., TCR sequences from~\cite{10x2019new} which were annotated as being paired with HLA-A*03:01 or HLA-A*11:01. We deduplicated the remaining data by their paired-chain TCR sequence; the annotated antigen species, epitope, and gene; and the annotated HLAs. A sequence from the INCOV {\trb} repertoires was matched to a sequence in the VDJdb if it was within 24 TCRdist of a VDJdb sequence and there was at least one identical HLA allele group between the INCOV individual from which the TCR originated and the VDJdb annotation.

\subsection*{Matching into the MIRA database}
The COVID-19 MIRA database (accessed November 14, 2024) was downloaded from Adaptive Biotechnologies COVID-2020 repository~\cite{ Nolan2025-fu}.
{\trb} sequences in the database underwent preprocessing in which gene names were conformed to IMGT conventions, and sequences were removed if they were unproductive or missing their  V or  J genes.
Moreover, the database was deduplicated at the level of   CDR3 nucleotide sequence,  V gene,  J gene, and individual.
A sequence from the INCOV {\trb} repertoires was said to be matched to the MIRA database, and thus specific for SARS-CoV-2, if it was within 12 TCRdist of at least two sequences in the MIRA database.

%

\clearpage{}
\newpage{}

\counterwithin{table}{section}
\setcounter{table}{0}
\renewcommand{\thetable}{S\arabic{table}}

\begin{table}
\begin{tabular}{|l|r|r|r|r|r|r|}
\hline
& \multicolumn{2}{c|}{\textbf{Healthy vs. PASC$^+$}} & \multicolumn{2}{c|}{\textbf{Healthy vs. PASC$^-$}} & \multicolumn{2}{c|}{\textbf{PASC$^+$ vs. PASC$^-$}} \\
\cline{2-7}
\textbf{V gene} & t statistic & p value & t statistic & p value & t statistic & p value \\ \hline
TRBV2 & -11.9 & $6.5 \times 10^{-27}$ & -10.2 & $3.9 \times 10^{-20}$ & -1.5 & $1 \times 10^{0}$ \\ \hline
TRBV3-1 & 79.7 & $6.7 \times 10^{-118}$ & 64.2 & $1.2 \times 10^{-80}$ & -0.6 & $1 \times 10^{0}$ \\ \hline
TRBV4-1 & -17.0 & $1.1 \times 10^{-48}$ & -17.2 & $2.8 \times 10^{-47}$ & 0.9 & $1 \times 10^{0}$ \\ \hline
TRBV4-2 & -13.4 & $6.8 \times 10^{-33}$ & -13.9 & $1.4 \times 10^{-34}$ & 1.2 & $1 \times 10^{0}$ \\ \hline
TRBV4-3 & -8.6 & $1.5 \times 10^{-14}$ & -9.2 & $1.6 \times 10^{-16}$ & 1.7 & $1 \times 10^{0}$ \\ \hline
TRBV5-1 & -6.3 & $6.3 \times 10^{-8}$ & -2.9 & $2.8 \times 10^{-1}$ & -2.5 & $4.6 \times 10^{-1}$ \\ \hline
TRBV5-4 & -11.0 & $6.5 \times 10^{-23}$ & -11.3 & $3.9 \times 10^{-24}$ & 1.3 & $1 \times 10^{0}$ \\ \hline
TRBV5-5 & -7.0 & $1.1 \times 10^{-9}$ & -7.0 & $1.1 \times 10^{-9}$ & 0.5 & $1 \times 10^{0}$ \\ \hline
TRBV5-6 & -16.0 & $6.9 \times 10^{-42}$ & -15.8 & $2.6 \times 10^{-41}$ & -0.2 & $1 \times 10^{0}$ \\ \hline
TRBV5-8 & -9.1 & $1.3 \times 10^{-15}$ & -9.6 & $2.3 \times 10^{-16}$ & 0.9 & $1 \times 10^{0}$ \\ \hline
TRBV6-1 & -24.4 & $1.7 \times 10^{-78}$ & -23.2 & $3.2 \times 10^{-74}$ & -1.8 & $1 \times 10^{0}$ \\ \hline
TRBV6-2 & NaN & NaN& NaN & NaN& 2.8 & $2.2 \times 10^{-1}$ \\ \hline
TRBV6-4 & -13.8 & $4.6 \times 10^{-33}$ & -14.2 & $1.0 \times 10^{-34}$ & 1.6 & $1 \times 10^{0}$ \\ \hline
TRBV6-5 & -6.6 & $7.3 \times 10^{-9}$ & -5.8 & $7.3 \times 10^{-7}$ & -0.7 & $1 \times 10^{0}$ \\ \hline
TRBV6-6 & -41.6 & $2.7 \times 10^{-156}$ & -39.0 & $1.2 \times 10^{-137}$ & -0.3 & $1 \times 10^{0}$ \\ \hline
TRBV7-2 & 4.2 & $2.1 \times 10^{-3}$ & 1.5 & $1 \times 10^{0}$ & 2.2 & $1 \times 10^{0}$ \\ \hline
TRBV7-3 & 1.1 & $1 \times 10^{0}$ & 0.5 & $1 \times 10^{0}$ & 0.5 & $1 \times 10^{0}$ \\ \hline
TRBV7-6 & -4.3 & $1.8 \times 10^{-3}$ & -4.2 & $2.6 \times 10^{-3}$ & 0.0 & $1 \times 10^{0}$ \\ \hline
TRBV7-8 & -7.2 & $2.3 \times 10^{-10}$ & -6.7 & $4.4 \times 10^{-9}$ & -0.4 & $1 \times 10^{0}$ \\ \hline
TRBV7-9 & -17.2 & $8.9 \times 10^{-50}$ & -18.8 & $3.2 \times 10^{-55}$ & 1.1 & $1 \times 10^{0}$ \\ \hline
TRBV9 & 12.2 & $1.2 \times 10^{-27}$ & 7.8 & $3.0 \times 10^{-11}$ & 1.1 & $1 \times 10^{0}$ \\ \hline
TRBV10-1 & -8.0 & $1.3 \times 10^{-12}$ & -7.4 & $1.8 \times 10^{-10}$ & -0.1 & $1 \times 10^{0}$ \\ \hline
TRBV10-2 & -6.7 & $8.2 \times 10^{-9}$ & -5.2 & $3.6 \times 10^{-5}$ & -1.7 & $1 \times 10^{0}$ \\ \hline
TRBV10-3 & -10.6 & $1.3 \times 10^{-21}$ & -9.7 & $1.8 \times 10^{-18}$ & -0.6 & $1 \times 10^{0}$ \\ \hline
TRBV11-2 & 2.3 & $1 \times 10^{0}$ & 2.9 & $3.3 \times 10^{-1}$ & -0.9 & $1 \times 10^{0}$ \\ \hline
TRBV11-3 & 3.2 & $9.5 \times 10^{-2}$ & 1.2 & $1 \times 10^{0}$ & 1.8 & $1 \times 10^{0}$ \\ \hline
TRBV12-3 & NaN & NaN& NaN & NaN& 0.7 & $1 \times 10^{0}$ \\ \hline
TRBV12-5 & -0.4 & $1 \times 10^{0}$ & -0.6 & $1 \times 10^{0}$ & 0.4 & $1 \times 10^{0}$ \\ \hline
TRBV13 & -6.2 & $1.2 \times 10^{-7}$ & -5.1 & $3.6 \times 10^{-5}$ & -1.6 & $1 \times 10^{0}$ \\ \hline
TRBV14 & -0.4 & $1 \times 10^{0}$ & -1.6 & $1 \times 10^{0}$ & 1.9 & $1 \times 10^{0}$ \\ \hline
TRBV18 & 8.0 & $5.9 \times 10^{-13}$ & 7.9 & $6.4 \times 10^{-12}$ & -1.5 & $1 \times 10^{0}$ \\ \hline
TRBV19 & 2.3 & $1 \times 10^{0}$ & 1.9 & $1 \times 10^{0}$ & 0.5 & $1 \times 10^{0}$ \\ \hline
TRBV20-1 & 99.4 & $2.1 \times 10^{-135}$ & 81.8 & $1.2 \times 10^{-92}$ & -2.5 & $4.8 \times 10^{-1}$ \\ \hline
TRBV24-1 & NaN & NaN& NaN & NaN& 1.6 & $1 \times 10^{0}$ \\ \hline
TRBV25-1 & -10.4 & $2.0 \times 10^{-20}$ & -10.1 & $1.1 \times 10^{-19}$ & -0.2 & $1 \times 10^{0}$ \\ \hline
TRBV27 & 1.0 & $1 \times 10^{0}$ & 1.6 & $1 \times 10^{0}$ & -0.9 & $1 \times 10^{0}$ \\ \hline
TRBV28 & 0.2 & $1 \times 10^{0}$ & 1.6 & $1 \times 10^{0}$ & -1.3 & $1 \times 10^{0}$ \\ \hline
TRBV29-1 & -4.3 & $1.3 \times 10^{-3}$ & -3.7 & $1.7 \times 10^{-2}$ & -0.4 & $1 \times 10^{0}$ \\ \hline
TRBV30 & 0.0 & $1 \times 10^{0}$ & -0.7 & $1 \times 10^{0}$ & 0.8 & $1 \times 10^{0}$ \\ \hline
\end{tabular}
\caption{Differential TRBV-gene usages among PASC{$^+$}, PASC{$^-$},  CMV{$^-$} healthy control cohorts are shown as t-statistics. The reported p-values are Bonferroni corrected. 
\label{table:vgene_ttests}}
\end{table}

\begin{table}
\begin{tabular}{|l|r|r|r|r|r|r|}
\hline
& \multicolumn{2}{c|}{\textbf{Healthy vs. PASC$^+$}} & \multicolumn{2}{c|}{\textbf{Healthy vs. PASC$^-$}} & \multicolumn{2}{c|}{\textbf{PASC$^+$ vs. PASC$^-$}} \\
\cline{2-7}
J gene & t statistic & p value & t statistic & p value & t statistic & p value\\ \hline
TRBJ1-1 & -13.2 & $5.0 \times 10^{-33}$ & -13.3 & $1.3 \times 10^{-32}$ & 0.7 & $1 \times 10^{0}$ \\ \hline
TRBJ1-2 & 4.4 & $4.0 \times 10^{-4}$ & 4.8 & $6.5 \times 10^{-5}$ & -0.7 & $1 \times 10^{0}$ \\ \hline
TRBJ1-3 & 0.3 & $1 \times 10^{0}$ & 0.0 & $1 \times 10^{0}$ & 0.6 & $1 \times 10^{0}$ \\ \hline
TRBJ1-4 & -10.8 & $7.2 \times 10^{-23}$ & -10.5 & $7.7 \times 10^{-22}$ & 0.0 & $1 \times 10^{0}$ \\ \hline
TRBJ1-5 & 0.0 & $1 \times 10^{0}$ & -0.2 & $1 \times 10^{0}$ & 0.3 & $1 \times 10^{0}$ \\ \hline
TRBJ1-6 & 14.4 & $3.0 \times 10^{-38}$ & 12.8 & $3.2 \times 10^{-30}$ & 1.6 & $1 \times 10^{0}$ \\ \hline
TRBJ2-1 & 0.8 & $1 \times 10^{0}$ & 1.7 & $1 \times 10^{0}$ & -1.2 & $1 \times 10^{0}$ \\ \hline
TRBJ2-2 & 4.2 & $7.4 \times 10^{-4}$ & 2.8 & $1.4 \times 10^{-1}$ & 1.7 & $1 \times 10^{0}$ \\ \hline
TRBJ2-3 & 5.5 & $1.4 \times 10^{-6}$ & 4.8 & $5.0 \times 10^{-5}$ & 0.6 & $1 \times 10^{0}$ \\ \hline
TRBJ2-4 & 2.1 & $8.6 \times 10^{-1}$ & 1.5 & $1 \times 10^{0}$ & 1.4 & $1 \times 10^{0}$ \\ \hline
TRBJ2-5 & -2.2 & $7.3 \times 10^{-1}$ & -2.1 & $1 \times 10^{0}$ & -0.5 & $1 \times 10^{0}$ \\ \hline
TRBJ2-6 & -13.3 & $4.4 \times 10^{-32}$ & -13.8 & $2.5 \times 10^{-34}$ & 1.7 & $1 \times 10^{0}$ \\ \hline
TRBJ2-7 & -5.2 & $9.7 \times 10^{-6}$ & -4.4 & $4.9 \times 10^{-4}$ & -0.5 & $1 \times 10^{0}$ \\ \hline
\end{tabular}
\caption{Differential TRBJ-gene usages among PASC{$^+$}, PASC{$^-$},  CMV{$^-$} healthy control cohorts are shown as t-statistics. The reported p-values are Bonferroni corrected.
\label{table:jgene_ttests}}
\end{table}

\begin{table}
  \begin{tabular}{|l|r|r|} \hline
    Groups & t-statistic & p-value \\ \hline
    PASC$^+$ vs. PASC$^-$ & -2.1 & 0.117 \\ \hline
    PASC$^+$ vs. Healthy & 49.2 & $3.5 \times 10^{-197}$ \\\hline
    PASC$^-$ vs. Healthy & 47.6 & $5.8 \times 10^{-170}$ \\\hline
  \end{tabular}
  \caption{Differential mean CDR3 lengths among PASC{$^+$}, PASC{$^-$},  CMV{$^-$} healthy control cohorts are shown as t-statistics. The reported p-values are Bonferroni corrected. Mean CDR3 lengths are 15.0 amino acids in PASC{$^+$} and  PASC{$^-$} repertoires, and 14.4 amino acids in CMV{$^-$} healthy controls.
    \label{table: cdr3_ttests}}
\end{table}

\begin{table}[htbp]
\begin{tabular}{|c | c | c | c | c|} 
 \hline
 PASC status & dynamic & null repertoire & $Z$ statistics & Bonferroni-corrected $p$-value \\
  \hline
 PASC{$^+$} & contracted & most abundant & $10.02$ & $p = 4.99 \times 10^{-23}$\\
\hline
 PASC{$^+$} & contracted & random & $15.47$ & $p = 2.31 \times 10^{-53}$\\
  \hline
 PASC{$^+$} & expanded & most abundant & $1.84$ & $p = 0.27$\\
\hline
 PASC{$^+$} & expanded & random & $8.51$ & $p = 6.68 \times 10^{-17}$\\
 \hline
 PASC{$^-$} & contracted & most abundant & $13.43$ & $p = 1.69 \times 10^{-40}$\\
\hline
 PASC{$^-$} & contracted & random & $16.88$ & $p = 2.33 \times 10^{-63}$\\
 \hline
 PASC{$^-$} & expanded & most abundant & $-0.74$ & $p = 1.0$\\
\hline
 PASC{$^-$} & expanded & random & $4.53$ & $p = 2.38 \times 10^{-5}$ \\[1ex] 
\hline
\end{tabular}
\caption{The differential amount of overlap between the MIRA dataset and the contracted / expanded repertoire subset in PASC{$^+$} and  PASC{$^-$}  and the corresponding null subsets (random draws or most abundant clones) is quantified using Z-statistics. Reported p-values are Bonferroni-corrected.
\label{table:tcrcovid_mira_dynamics}}
\end{table}

\begin{table}[h]
\centering
\begin{tabular}{llccc|ccc} 
\toprule
 & & \multicolumn{3}{c}{{\bf contracted}} & \multicolumn{3}{c}{{\bf expanded}} \\
\cmidrule(lr){3-5} \cmidrule(lr){6-8}
 & & \makecell{No.\ of individuals \\ (PASC{$^+$} ; PASC{$^-$})} & MW statistic & $p$-value & \makecell{No.\ of individuals \\ (PASC{$^+$} ; PASC{$^-$})} & MW statistic & $p$-value \\
\hline
\multirow{2}{*}{Influenza A} & matching       & \multirow{2}{*}{(15 ; 7)}  & 86.5                  & \multirow{2}{*}{0.015} & \multirow{2}{*}{(12 ; 5)}  & 32.5                  & \multirow{2}{*}{0.593} \\
                             & null bootstrap &                         & $59.76 \pm 12.88$     &                        &                         & $34.10 \pm 8.55$      & \\
\hline
\multirow{2}{*}{CMV}         & matching       & \multirow{2}{*}{(11 ; 9)}  & 80                    & \multirow{2}{*}{0.024} & \multirow{2}{*}{(15 ; 11)} & 115.5                 & \multirow{2}{*}{0.117} \\
                             & null bootstrap &                         & $56.32 \pm 12.27$     &                        &                         & $93.98 \pm 17.88$     & \\
\hline
\multirow{2}{*}{EBV}         & matching       & \multirow{2}{*}{(11 ; 6)}  & 47.5                  & \multirow{2}{*}{0.147} & \multirow{2}{*}{(10 ; 5)}  & 39                    & \multirow{2}{*}{0.081} \\
                             & null bootstrap &                         & $37.57 \pm 9.07$      &                        &                         & $28.42 \pm 7.40$      & \\
\hline
\multirow{2}{*}{SARS-CoV-2}  & matching       & \multirow{2}{*}{(19 ; 13)} & 166                   & \multirow{2}{*}{0.15}  & \multirow{2}{*}{(16 ; 6)}  & 60                    & \multirow{2}{*}{0.348} \\
                             & null bootstrap &                         & $140.48 \pm 24.16$    &                        &                         & $54.55 \pm 12.20$     & \\
\bottomrule
\end{tabular}
\caption{For contracted and expanded TCR clusters matching VDJ database antigen-specific TCRs in Fig.~\ref{fig:clonal_expansion_OT}E, we identified the individuals from whom these TCRs were derived. The number of such individuals in the PASC{$^+$} and  PASC{$^-$} groups is indicated in each case. We then computed the Mann-Whitney test statistic on the WHO ordinal severity scale, comparing PASC{$^+$} against  PASC{$^-$} individuals within each antigen category. To assess statistical significance, we generated a null distribution by drawing 100,000 random pairs of PASC{$^+$} and  PASC{$^-$} subsets of the same size from the full dataset, and computed the Mann-Whitney test statistic for each pair; the mean and the standard deviation of this test statistics across the 100,000  are reported. We then  derived a bootstrap $p$-value for the observed test statistic against this null distribution.
\label{table:SI_VDJDB_WHO} }
\end{table}

\clearpage
\newpage{}

\counterwithin{figure}{section}
\setcounter{figure}{0}
\renewcommand{\thefigure}{S\arabic{figure}}

\begin{figure*}[htbp]
    \makebox[\textwidth][c]{\includegraphics[width=\textwidth]{SI_repertoire_wide_seq_stats.pdf}}
    \centering
    \caption[Receptor composition at the level of unique recombinations across cohorts.]{
    {\bf Receptor composition at the level of unique recombinations across cohorts.}
    {\bf(A)} The distribution of  V gene usages (left),   J gene usages (middle), and $\beta$CDR3 lengths are shown for healthy individuals (from CMV{$^-$} Emerson cohort from ref.~\cite{emerson2017immunosequencing}), and PASC{$^+$} and PASC{$^-$}  repertoires (colors). Interquartiles are depicted in the boxplots, with the median shown as a black, horizontal line.
    The rest of the distribution is shown as whiskers, and outliers are plotted as black dots.
  {\bf (B-D)} Similar to (A) but  distributions stratified by PASC status and sex (colors) and sampling time relative to symptom onset as indicated above each triple of plots; note the overlapping histograms in  panels on the right. Panel D indicates the gene identities for corresponding columns in B and C. 
    {\bf (E)} The distribution of clonalities of the  TCR repertoires  (left) and the associated log-fold changes in clonality between time points  (right) are shown stratified by PASC status and sex (colors) as well as by sampling time relative to symptom onset (x-axis).
   On the left, the bars indicate the clonality averaged over individuals in each PASC$^-$sex group, and dots indicate the variation in clonalities across individuals within each PASC$^-$sex group. On the right, the medians of the distributions are demarcated by a horizontal black line.   For log-fold changes the x-axis denotes the numerator of the fold change and the denominator of the fold change as ``numerator vs. denominator."
    \label{fig:SI:tcrcovid_composition}
    }
\end{figure*}

\begin{figure*}[htbp]
{\includegraphics[width=0.7\textwidth]{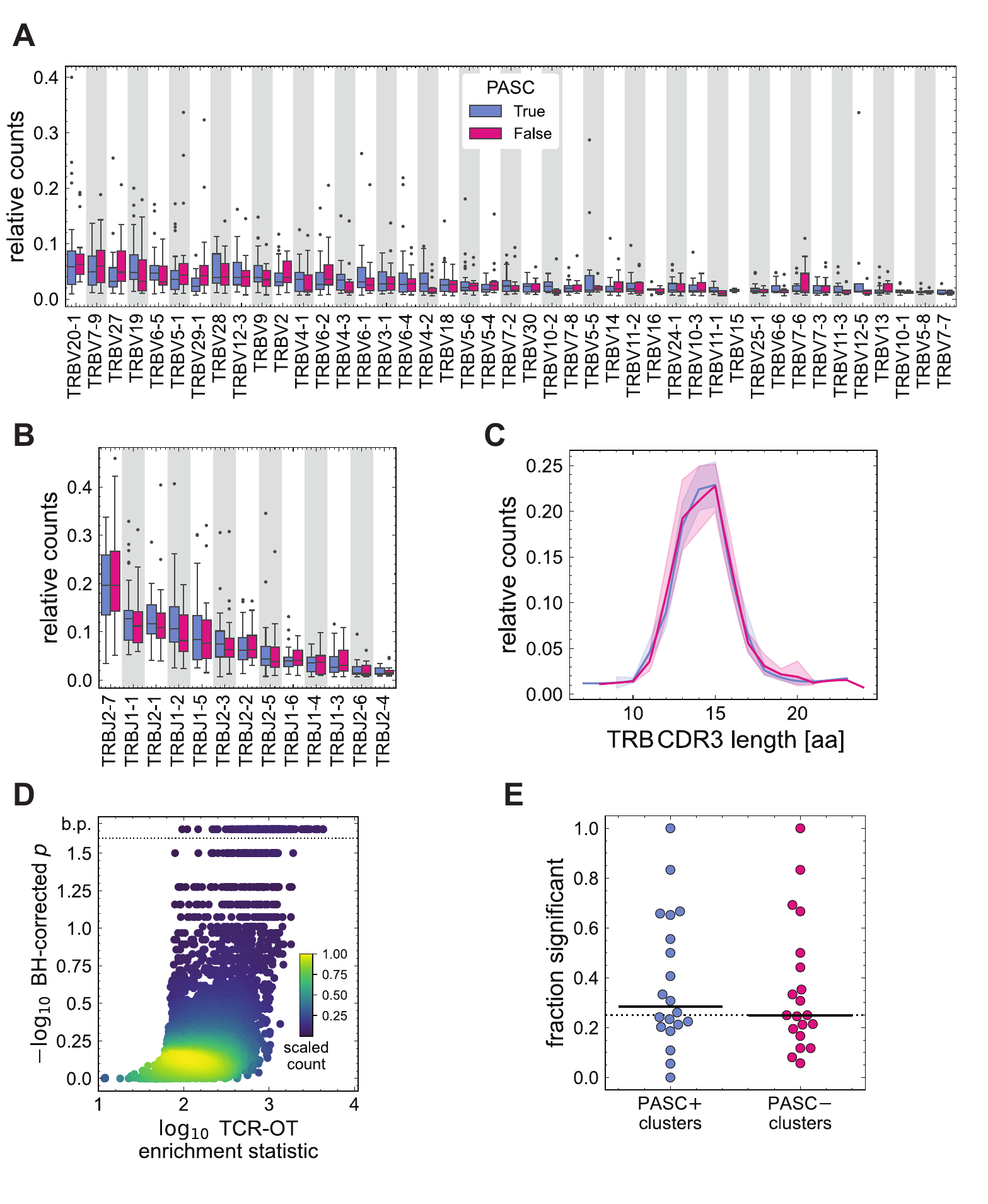}}
    \caption[Top rank-50 receptor composition and TCR-OT clusters.]{
    {\bf Top rank-50 receptor composition and TCR-OT clusters.}
    {\bf(A)} The distribution of  V gene usages is shown as boxplots, stratified by PASC status (colors).
    Interquartiles are depicted in the box, with the median shown as a black, horizontal line.
    The rest of the distribution is shown as whiskers, and outliers are plotted as black dots.
    {\bf(B)} The distribution of  J gene usages is shown as boxplots as in (A).
    {\bf(C)} The distribution of amino acid   CDR3 lengths is shown.
    Lines indicate average relative counts for each PASC group, and shading indicates regions containing one standard deviation of variation across individuals within a group.\
    {\bf(D)} The volcano plot shows the $\log_{10}$ FDR BH-corrected $p$-value vs. the $\log_{10}$ TCR-OT enrichment statistic for each functional clone tested.
    Points above the dotted line, annotated along the y-axis as having a value of b.p. were assigned $p$-values beyond the precision of the randomization test, i.e., a value of 0.
    {\bf(E)} The swarm plot shows the distribution of the fraction of functional clones which were deemed significant within a TCR-OT cluster for each group (colors).
    The black horizontal lines show the median for each distribution.
    Markers above the horizontal, dotted line indicate clusters which have more than 20\% of their constituent clones called as significant.
       \label{fig:SI:tcrcovid_tcrot_top50_composition}
    }
\end{figure*}

\begin{figure*}[htbp]
{\includegraphics[width=0.8\textwidth]{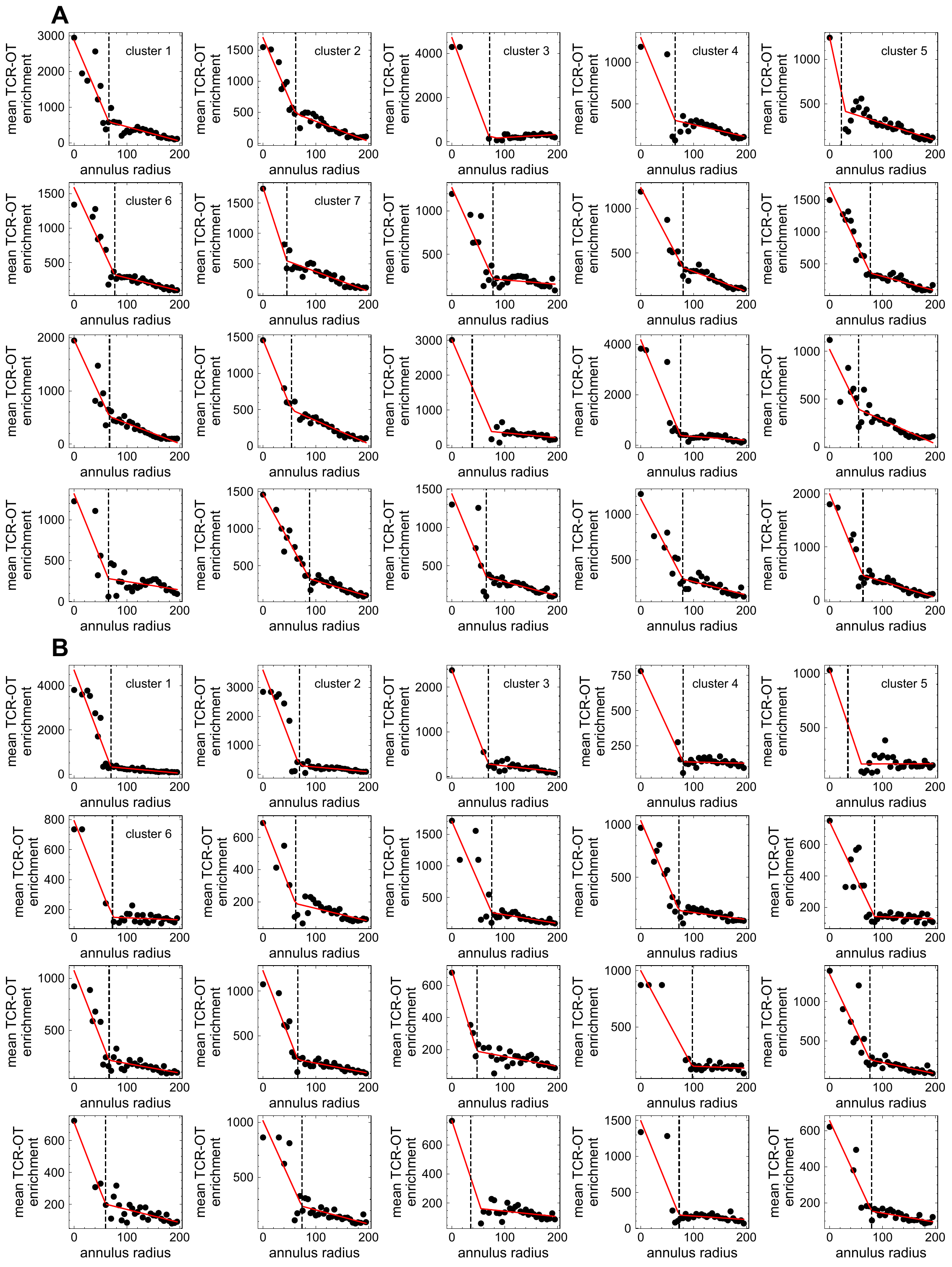}}
  \caption[TCR-OT cluster breakpoint analyses on top rank-50 repertoires]{{\bf TCR-OT cluster breakpoint analyses on top rank-50 repertoires}
    The mean TCR-OT enrichment vs annulus radius (given in units TCRdist) for cluster detection in the PASC{$^+$} top 50-rank repertoire {\bf(A)} and PASC{$^-$} top 50-rank repertoire {\bf(B)}.
    The black dots show the observed statistic at within the annulus, and the red line shows the segmented linear fit (Methods).
    The breakpoint, shown as a vertical black, dashed line, in each segmented linear regression determines the cluster radius.
    Regressions associated with clusters used in downstream analyses are annotated with their cluster size rank, as in Fig.~\ref{fig:tcrcovid_tcrot_composition}.
    \label{fig:tcrcovid_tcrot_top50_clusterfit}
}
\end{figure*}

\begin{figure}[htbp]
    \makebox[\textwidth][c]{\includegraphics[width=0.9\textwidth]{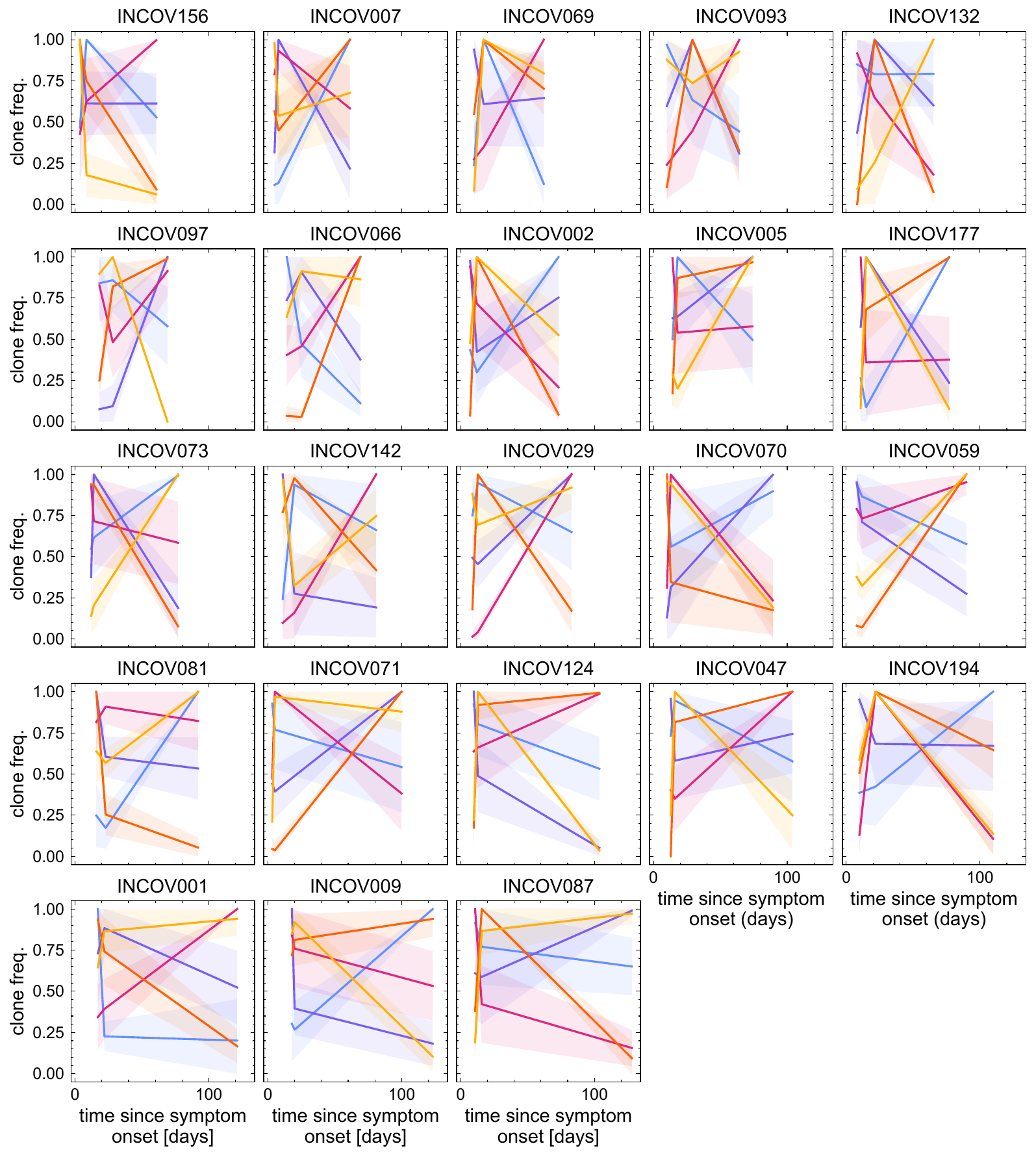}}
    \centering
\caption[Dynamical modes of clonal frequency trajectories for PASC$^+$ cohort.]{
     {\bf Dynamical modes of clonal frequency trajectories for PASC{$^+$} repertoires.}
    The modes of clonal trajectories from the top rank-1000 clones within an individual from the  PASC$^+$ group, using PCA and hierarchical clustering.
    Lines indicate average clonal frequency within each mode (colors), and shading indicates regions containing one standard deviation of variation across the mode.
    Plots are ordered from left-to-right and top-to-bottom using the time at which the individual's repertoire was last sampled.
    \label{fig:SI_tcrcovid_pasc_dynamic_modes}
    }
\end{figure}

\begin{figure}[htbp]
    \makebox[\textwidth][c]{\includegraphics[width=0.9\textwidth]{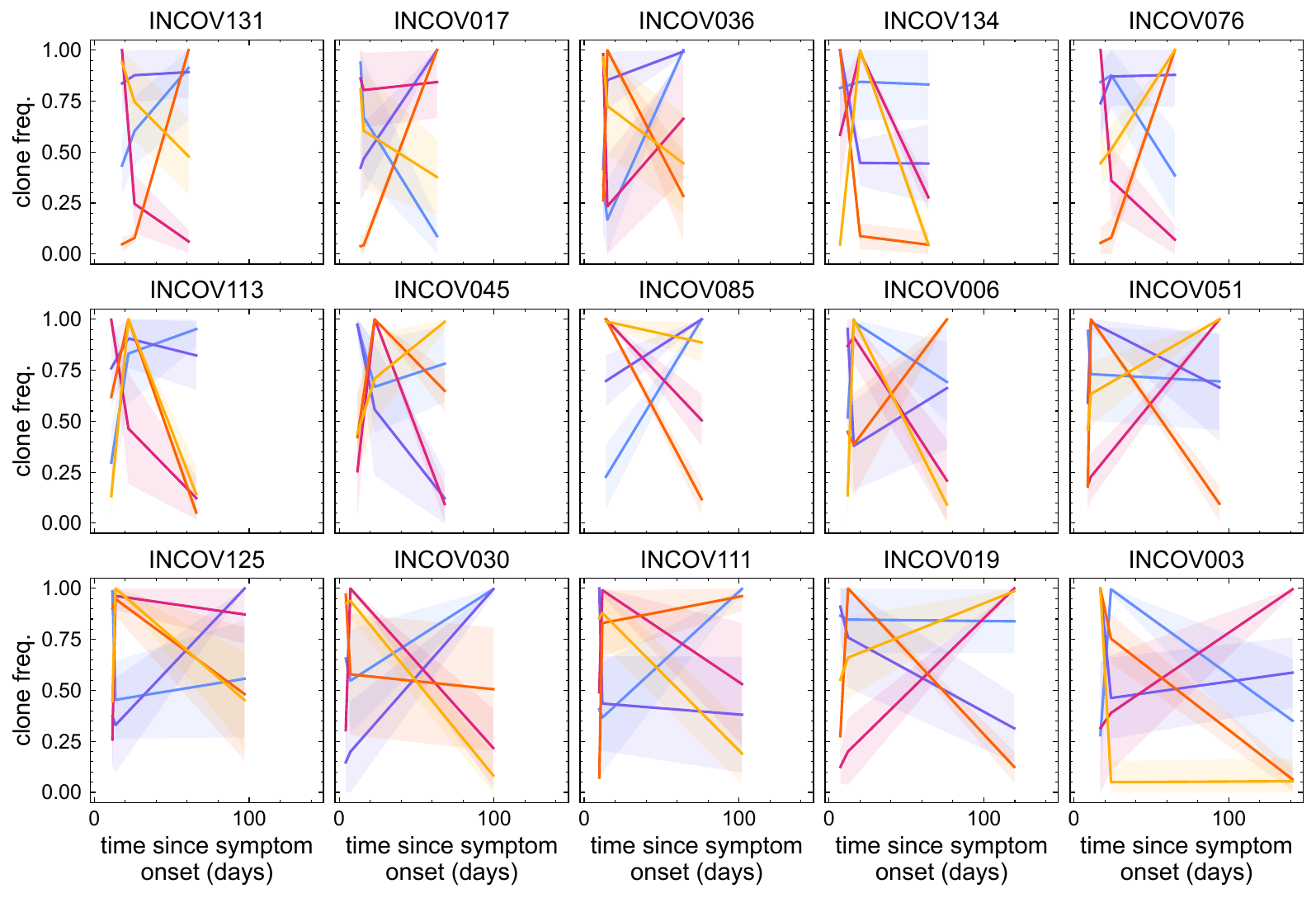}}
    \centering
    \caption[Dynamical modes of clonal frequency trajectories for PASC{$^-$} cohort.]{
    {\bf Dynamical modes of clonal frequency trajectories for PASC{$^-$} repertoires.}
    Similar to  Fig.~\ref{fig:SI_tcrcovid_pasc_dynamic_modes} but for individuals from the   PASC{$^-$} group.
    \label{fig:SI_tcrcovid_nonpasc_dynamic_modes}
    }
\end{figure}

\begin{figure}[htbp]
    \makebox[\textwidth][c]{\includegraphics[width=0.7\textwidth]{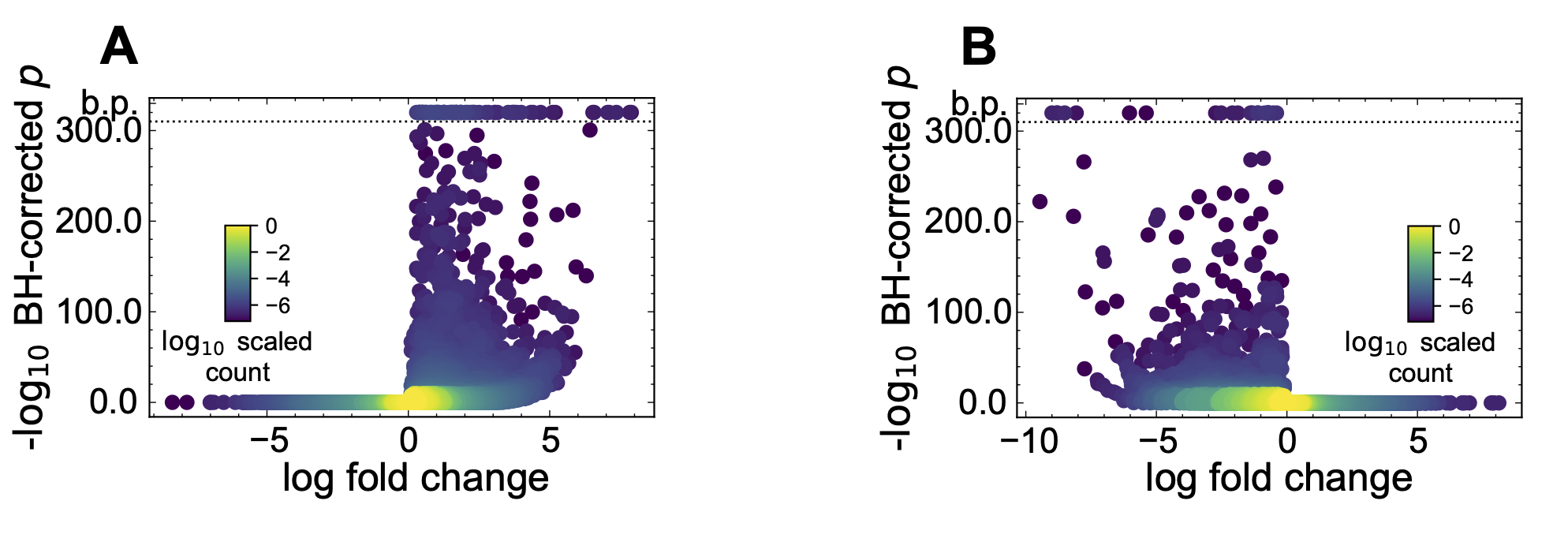}}
    \caption[NoisET volcano plots]{
    {\bf NoisET volcano plots for expanding and contracting TCR clones.}
    {\bf(A)} The volcano plot shows the $-\log_{10}$ BH-corrected $p$-value vs. the NoisET-inferred mean log-fold change for each functional clone tested for expansion.
    $p$-values annotated as b.p. are beyond machine precision.
    {\bf(B)} As in (A) but for each functional clone tested for contraction.
    \label{fig:SI_dynamics_volcanos_noiset}
    }
\end{figure}

\begin{figure}[htbp]
    \makebox[\textwidth][c]{\includegraphics[width=1.\textwidth]{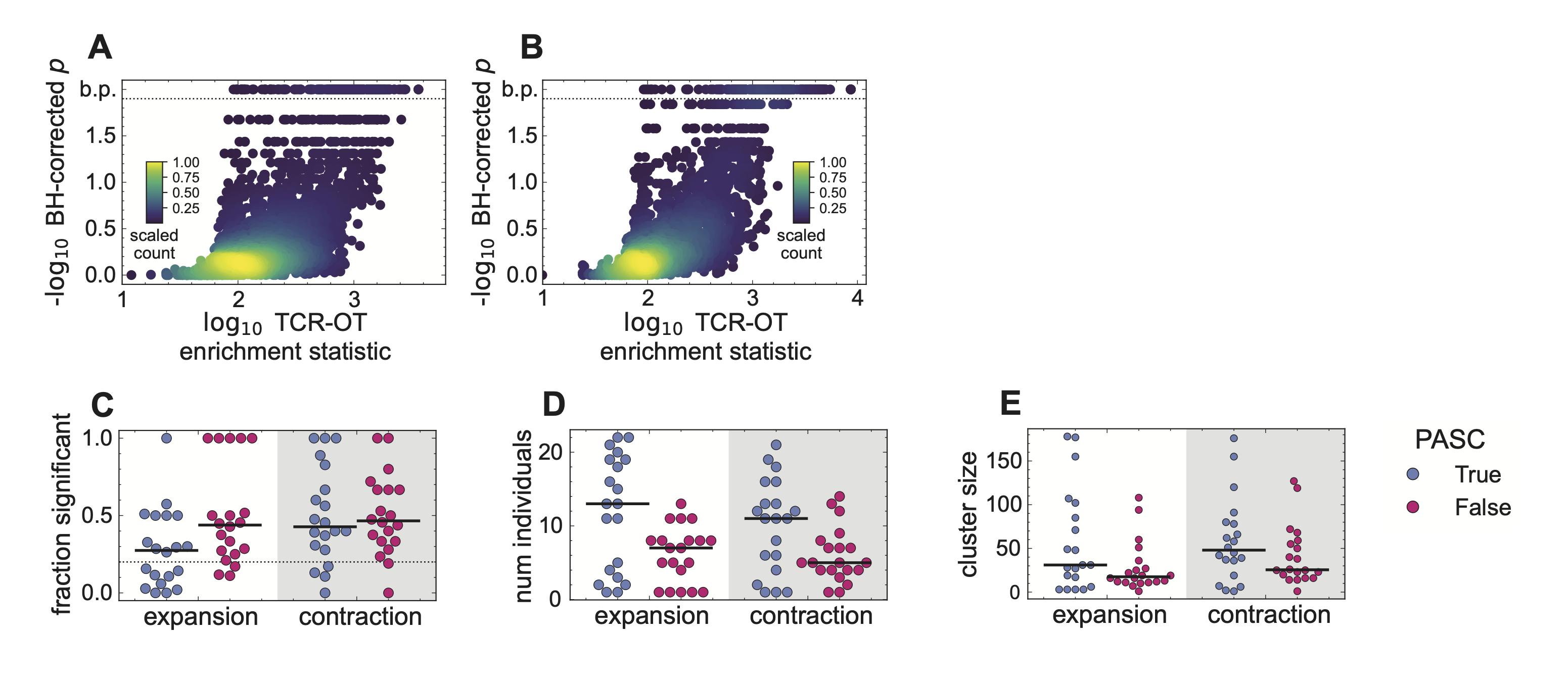}}
    \caption[TCR-OT summary statistics and clusters.]{
    {\bf TCR-OT summary statistics for dynamic clones.}
        {\bf(A)} The volcano plot shows the $-\log_{10}$ BH-corrected $p$-value vs. the $\log_{10}$ TCR-OT enrichment statistic for each functional clone in the expanded PASC{$^+$} and PASC{$^-$} repertoires.
    $p$-values annotated as b.p. are beyond machine precision.
  {\bf(B)} As in (A) but for each functional clone in the contracted PASC{$^+$} and PASC{$^-$} repertoires.
  {\bf(C)} The swarm plot shows the distribution of the fraction of functional clones which were deemed significant within each of the 20  TCR-OT clusters (Figs.~\ref{fig:SI_tcrcovid_tcrot_expanding_clusterfit},~\ref{fig:SI_tcrcovid_tcrot_contracting_clusterfit}) stratified by dynamic (x-axis) and PASC status (colors).
  The black horizontal lines show the median for each distribution.
  Markers above the horizontal, dotted line indicate clusters which have more than 20\% of their constituent clones called as significant.
  {\bf(D)} The swarm plot shows the distribution of number of individuals in each TCR-OT cluster stratified by dynamic (x-axis) and PASC status (colors).
  {\bf(E)} The swarm plot shows the  distribution for the number of  independent recombinations (cluster size) in each TCR-OT cluster stratified by dynamic (x-axis) and PASC status (colors).
    \label{fig:SI_dynamics_volcanos_ot}
    }
\end{figure}

\begin{figure*}[htbp]
{\includegraphics[width=0.8\textwidth]{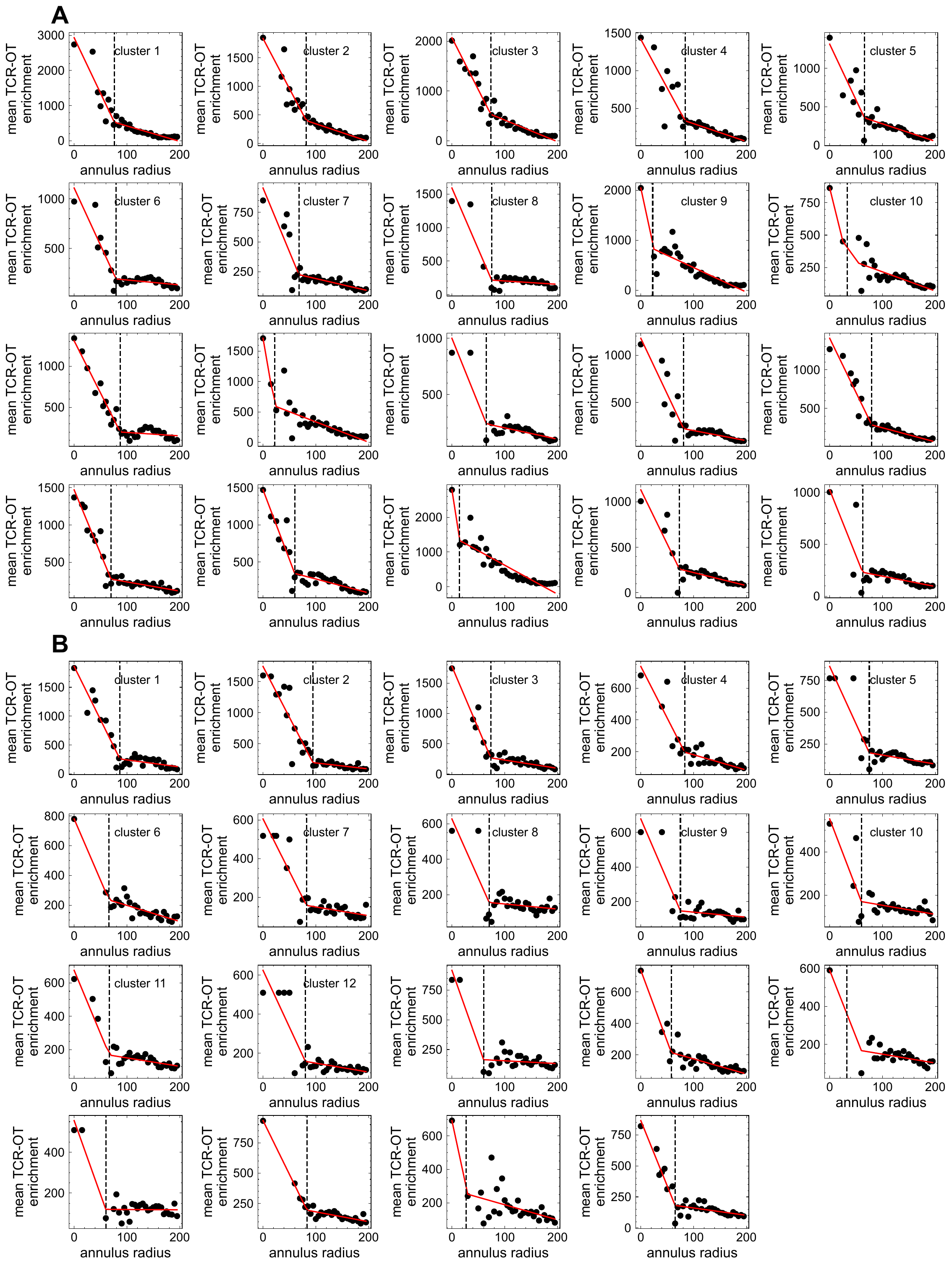}}
  \caption{{\bf TCR-OT cluster breakpoint analyses on the expanding clones}
  Similar to Fig.~\ref{fig:tcrcovid_tcrot_top50_clusterfit} but for significantly expanding clones identified in Fig.~\ref{fig:clonal_expansion}.
    \label{fig:SI_tcrcovid_tcrot_expanding_clusterfit}
}
\end{figure*}

\begin{figure*}[htbp]
{\includegraphics[width=0.8\textwidth]{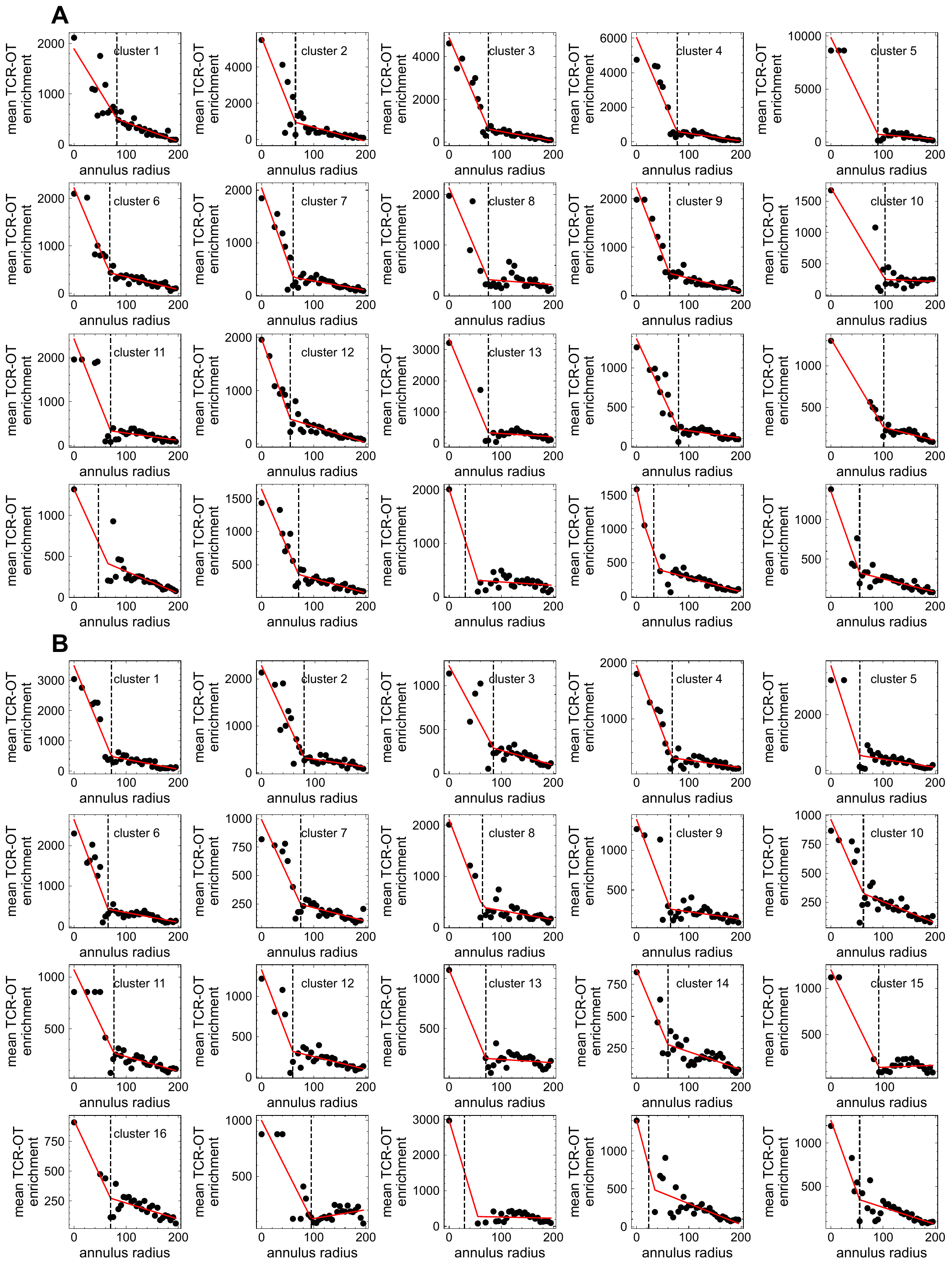}}
  \caption{{\bf TCR-OT cluster breakpoint analyses on the contracting clones.}
  Similar to Fig.~\ref{fig:tcrcovid_tcrot_top50_clusterfit} but for significantly contracting clones identified in Fig.~\ref{fig:clonal_expansion}.
    \label{fig:SI_tcrcovid_tcrot_contracting_clusterfit}
}
\end{figure*}

\begin{figure*}[htbp]
    \makebox[\textwidth][c]{\includegraphics[width=1\textwidth]{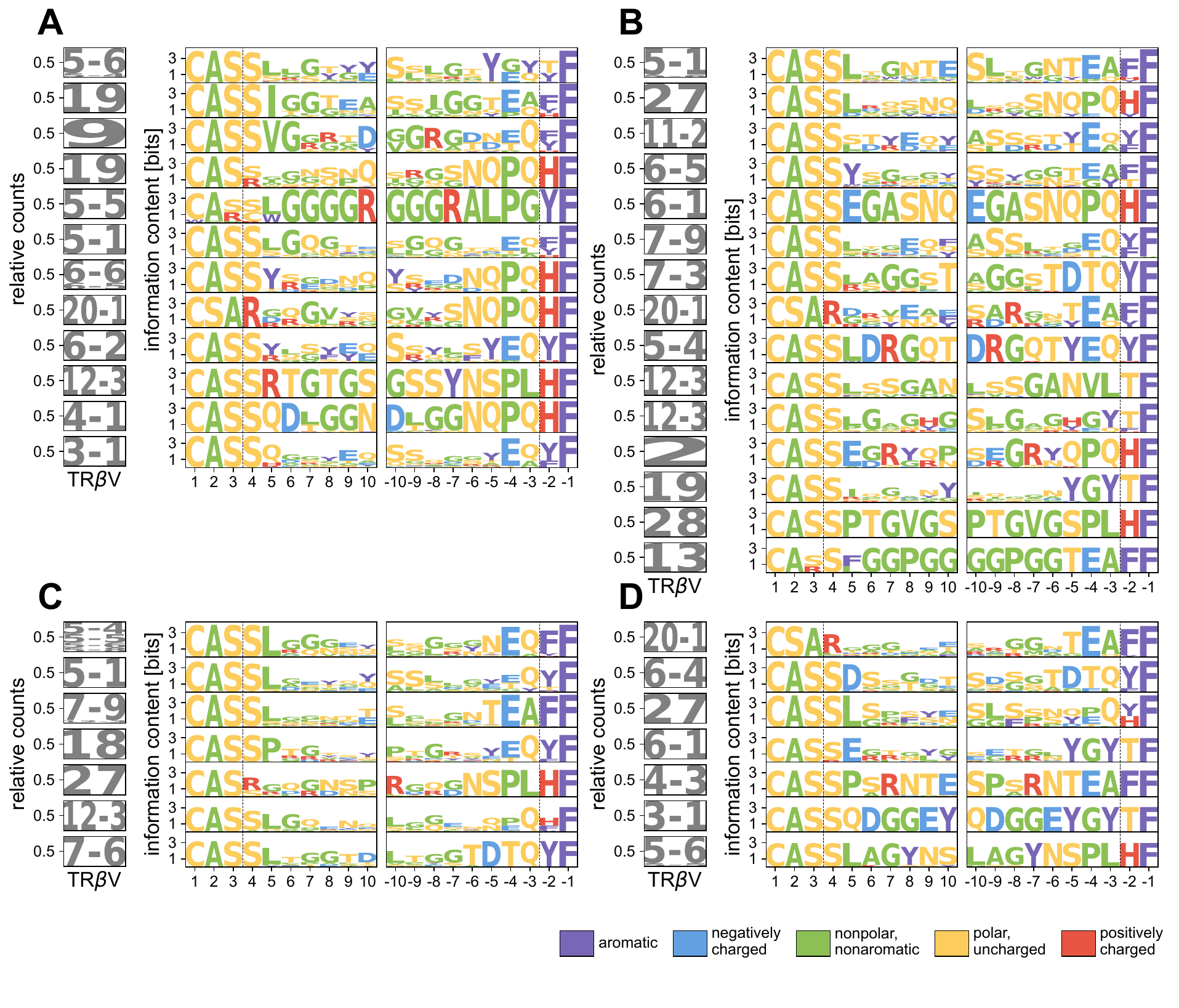}}
    \centering
  \caption[Motifs and VDJdb matching among dynamical TCR-OT clusters.]{
  {\bf Motifs and VDJdb matching among dynamical TCR-OT clusters.}
  {\bf(A, B, C, D)} Sequence logos describing the composition of amino acid sequences are shown for the significant TCR-OT clusters obtained from the contracted PASC{$^+$} subset (A), contracted PASC{$^-$} subset (B), expanded PASC{$^+$} subset (C), and expanded PASC{$^-$} subset (D).
    Logos were constructed from focal sequence(s) and members in the cluster that were within 48 TCRdist of the focal sequence(s).
    Logos are shown only if {at least} 10 sequences were present in this 48 TCRdist neighborhood.
    Logos are ordered top to bottom by cluster size, with the largest cluster size at the top.
    (Left)  V gene usage frequencies within a cluster are shown as a logo.
    (Middle, right) The logo plots shown the information content at each position in the amino acid {\tcrb} CDR3 sequence from left to right (middle) and right to left (right).
    Amino acids are colored according to their physicochemical properties indicated in the legend.
    \label{fig:SI_tcrcovid_tcrot_dynamics_motifs}
    }
\end{figure*}

\clearpage{}

\noindent {{\bf Dataset 1:} Cluster membership for different HLA alleles based on the allele's peptide binding cleft sequences are shown.}

\end{document}